\documentclass[fleqn,usenatbib]{mnras}

\usepackage{newtxtext,newtxmath}

\usepackage[T1]{fontenc}

\DeclareRobustCommand{\VAN}[3]{#2}
\let\VANthebibliography\thebibliography
\def\thebibliography{\DeclareRobustCommand{\VAN}[3]{##3}\VANthebibliography}

\usepackage{graphicx}	
\usepackage{amsmath}	

\newcommand{\Quijote}{\textsc{Quijote} }
\newcommand{\Mpc}{\rm{Mpc}}
\newcommand{\hMpc}{h^{-1} \Mpc}

\newcommand{\de}{\text{d}}

\title[CNN reconstruction]{Effective cosmic density field reconstruction with convolutional neural network}

\author[Chen et al.]{
Xinyi Chen$^{1}$\thanks{E-mail: xinyi.chen@yale.edu},
Fangzhou Zhu$^{2}$,
Sasha Gaines$^{3}$
and Nikhil Padmanabhan$^{1,3}$
\\
$^{1}$Department of Physics, Yale University, New Haven, CT 06511, USA\\
$^{2}$Google LLC\\
$^{3}$Department of Astronomy, Yale University, New Haven, CT 06511, USA\\
}

\date{Accepted XXX. Received YYY; in original form ZZZ}

\pubyear{2023}

\begin{document}
\label{firstpage}
\pagerange{\pageref{firstpage}--\pageref{lastpage}}
\maketitle

\begin{abstract}
We present a cosmic density field reconstruction method that augments the traditional reconstruction algorithms with a convolutional neural network (CNN).
Following Shallue \& Eisenstein (2022), the key component of our method is to use the \textit{reconstructed} density field as the input to the 
neural network. We extend this previous work by exploring how the performance of these reconstruction ideas depends on the input 
reconstruction algorithm, the reconstruction parameters, and the shot noise of the density field, as well as the robustness of the 
method.
We build an eight-layer CNN and train the network with reconstructed density fields computed from the \textsc{Quijote} suite 
of simulations. 
The reconstructed density fields are generated by both the standard algorithm and a new iterative algorithm. 
In real space at $z=0$, we find that the reconstructed field is 90\% correlated with the true initial density out to $k\sim0.5 h\Mpc^{-1}$, a significant improvement over $k\sim0.2 h\Mpc^{-1}$ achieved by the input reconstruction algorithms. 
We find similar improvements in redshift space, including an improved removal of redshift space 
distortions at small scales.
We also find that the method is robust across changes in cosmology. Additionally, the CNN removes much of the variance from the choice of different reconstruction algorithms and reconstruction parameters. However, the effectiveness decreases with increasing shot noise, suggesting that such 
an approach is best suited to high density samples. 
This work highlights the additional information in the density field beyond linear scales
as well as the power of complementing traditional analysis approaches with machine learning techniques.

\end{abstract}




\section{Introduction}
The baryon acoustic oscillation (BAO) technique is one of the most powerful probes of dark energy. The ``acoustic scale'' of about 150 $\Mpc$ provides a standard ruler to map the expansion history of the universe, which offers valuable information to study dark energy and evaluate different theories of dark energy, such as cosmological constant, modified gravity, and effects of quantum gravity. A precise measurement of the ``acoustic peak'' at the acoustic scale in a matter correlation function is critical to have success with the BAO technique. The acoustic peak can be used to constrain cosmic distance scales, the angular diameter distance and Hubble parameter at various redshift, thereby shedding light on dark energy. Ongoing and next generation large-scale structure surveys, such as Dark Energy Spectroscopic Instrument \citep[DESI,][]{DESI}, \textit{Nancy Grace Roman Space Telescope} \citep[][]{Spergel13} and \textit{Euclid} \citep[][]{Laureijs11}, all aim to achieve ever higher precision in BAO measurements.

One systematic that reduces the precision of measurement of BAO parameters in galaxy clustering is the nonlinear evolution of the late-time universe, such as cluster formation and structure growth. These effects show up in the matter correlation function as a dampened and broadened acoustic peak and in the power spectrum as erased higher harmonics \citep[e.g.,][]{Meiksin99,Eisenstein07b,Eisenstein07,Crocce08,padmanabhan09b,Seo10}. A way to mitigate this effect is by cosmic density field reconstruction, which attempts to reverse the large-scale bulk flows. The product of reconstruction of the density field can also benefit beyond measuring BAO and probing the nature of dark energy, such as full shape analysis, which uses the full shape of power spectrum to probe other aspects of cosmology, such as the geometry of the space \citep[AP effect,][]{Alcock79} and redshift space distortions \citep[RSD,][]{Kaiser87}. Efforts have been made in this front. For example, \citet[][]{White15} and \citet[][]{Chen19} model the post-reconstruction two-point statistics with Zel'dovich approximation. These allow a fit for AP effect and RSD as well as growth rate of initial perturbation, $f$, along with BAO. Reconstruction may also benefit primordial non-Gaussianity studies by removing gravitationally induced non-Gaussianities.

\citet{Eisenstein07} proposed the first BAO reconstruction method using first order Lagrangian perturbation theory \citep[Zel'dovich displacement,][]{Zeldovich70} to estimate the displacement of galaxies. This method has been used in observations for over a decade achieving improvement in precision of BAO measurements by a factor of 1.2-2.4 \citep[e.g.,][]{Padmanabhan12,Xu13,Anderson14b,Beutler17,Ross17,Vargas18,Gil20} and has been consequently referred to as the ``standard method''. Over the years, there have been a number of reconstruction methods proposed that reconstructs either the displacement or the density field \citep[e.g.,][]{Seo10,Tassev12,Schmittfull15,Seo16,Obuljen17,Schmittfull17,Hada18,Levy21,Nikakhtar22,Seo22}. The new components of these methods include employing an iterative process, using annealing smoothing scales, and introducing second-order perturbation theory or a different way of estimating the displacement altogether \citep[e.g., optimal transport,][]{Levy21,Nikakhtar22}, etc. However, these new methods generally perform as well as the standard method, although they may outperform in certain aspects, such as removing RSD \citep[][]{reconpaper}. 

There are also at least two algorithms that do not fall into the above category. There is a conceptually different way of reconstruction that does not start from the nonlinear density field, but samples the initial condition and then evolves the field with forward modeling \citep[e.g.,][]{Jasche13,Kitaura13,Wang14,Seljak17}. There is also a method to reconstruct the two-point correlation function directly using Laguerre functions and does not involve estimating the displacement or density field \citep[][]{Nikakhtar21}.

All of the above reconstruction algorithms are based on large-scale structure physics. While they achieve excellent recovery of the linear field, the approximation to the initial condition has the potential to be further improved and the large-scale structure surveys desire more improvement in reconstruction to more fully realize their capability of probing cosmology. With tremendous amount of simulation data and powerful enough computing resources available, it is necessary to explore more data-driven approaches for reconstruction now.

In the past decade, deep learning has become increasingly powerful with the arrival of huge amount of data and has been shown to outperform traditional machine learning methods that rely on hand-crafted features. One category of deep learning techniques excellent at recognizing image features is convolutional neural networks \citep[CNNs, e.g.,][]{Fukushima82,LeCun99,Krizhevsky12,Simonyan14}. CNNs employ a hierarchical structure, common in real-world images, to extract underlying and complex patterns, such as edges, textures and shapes. 

However, the ability of CNNs is not limited within image data. In cosmology, with the availability of tremendous amount of simulation data, CNNs have been applied for cosmological inference and accelerating cosmological simulations \citep[][]{Company22}. In terms of constraining cosmological parameters with large-scale structure, CNNs have been applied to 3D dark matter or galaxy distributions to estimate cosmological parameters and it has been shown that CNN can extract more cosmological information from 3D large-scale structure than power spectrum \citep[e.g.,][]{Ravanbakhsh17,Mathuriya18,Ntampaka20}. CNN has been more researched in weak gravitational lensing cosmology to constrain cosmological parameters, mostly ($\sigma_8$,$\Omega_{\rm m}$) \citep[e.g.,][]{Gupta18,Fluri18,Ribli19}. Moreover, CNN also helps generate cosmological simulations more rapidly, by learning $N$-body simulations, or $N$-body emulation with deep learning, etc. \citep[e.g.,][]{Dai18,Mustafa19}. These all have achieved promising results. Beyond cosmological inference and simulation acceleration, CNNs have been applied in multiple areas, e.g., to estimate cluster or halo masses \citep[e.g.,][]{Ntampaka19,Ho19,Lucie-Smith20,Etezad-Razavi21}. For more applications of CNN in cosmology, see the review by \citet[][]{Company22}.

In cosmic density reconstruction, neural networks, including CNNs, have been exploited as well, but studies are still at an exploratory level. \citet[][]{Modi18} propose a forward modeling based approach for reconstruction and a fully-connected neural network is applied to predict halo mass and position field from matter density field. \citet{Mao21} more directly use a CNN for reconstruction. They use dark matter simulations and obtain results better than the standard method in certain scales. Most recently, \citet[][]{Shallue22} propose a CNN model for reconstruction using reconstructed fields as input and achieve a factor of 2 improvement in cross-correlation.  

In this work, we build our CNN model modifying \citet{Mao21}, but train on \textit{reconstructed} density fields rather than raw, nonlinear fields. We employ two reconstruction methods and attempt further reconstruction with CNN in matter and shot noise fields and in both real and redshift space using dark matter simulations. We use propagator, cross-correlation coefficient, and power spectrum as metrics to quantify the performance of our method compared to that of reconstruction algorithms. Our work resembles \citet[][]{Shallue22} in terms of the principle, but our CNN architecture, simulations, and metrics for analysis are different. We also study the performance of CNN model that trains on density fields reconstructed from a new algorithm by \citet[][]{Hada18}, beyond standard reconstruction.

This paper is organized as follows: Section~\ref{sect:model} details our CNN architecture. Section~\ref{sect:recon} describes the two reconstruction methods used in this work. Section~\ref{sect:results} presents results and robustness, where we show the performance in real and redshift space at $z=0$, model applied to different cosmologies and redshift, and extension to high shot-noise fields. In Section~\ref{sect:discussion} we discuss our results and we conclude and discuss future work in Section~\ref{sect:summary}.

\section{Model}\label{sect:model}
Our model is an eight-layer CNN, with seven convolutional layers and one fully-connected layer. It is designed to predict the density at a central grid point based on a neighborhood of $39^3$ grid points. Given our standard grid spacing of 1.95 $\hMpc$, this corresponds to a region of $76.17^3(\hMpc)^3$.
The architecture is based on \citet{Mao21}, but we modify it to achieve better performance (see Table~\ref{tab:architecture} for more details). A minor disadvantage of this approach is that the network architecture is linked to the underlying grid. For example, if one halved the grid spacing, one would need to increase the number of convolutional layers, or to change the kernel size to use the same physical neighborhood, or use a strided kernel. We defer such explorations to future work.

\begin{table*}
\centering
\begin{tabular}{cccccc}
\hline
 & Layer  & Kernel size & Stride size & Dilation size & Output grid size $\times$ channels\\ \hline
 & (Input) & - & - & - & $39^3 \times 1$  \\
(1) & 3D convolution  & 3$\times$3$\times$3 & 1 & 1 & $37^3 \times 32$\\
(2) & 3D convolution  & 3$\times$3$\times$3 & 1 & 2 & $33^3 \times 32$\\
(3) & 3D convolution  & 3$\times$3$\times$3 & 1 & 2 & $29^3 \times 64$\\
(4) & 3D convolution  & 3$\times$3$\times$3 & 1 & 4 & $21^3 \times 64$\\
(5) & 3D convolution  & 3$\times$3$\times$3 & 1 & 4 & $13^3 \times 128$\\
(6) & 3D convolution  & 2$\times$2$\times$2 & 1 & 8 & $5^3 \times 128$ \\
(7) & 3D convolution  & 1$\times$1$\times$1 & 1 & 1 & $5^3\times 128$ \\
& - & - & - & - & trimmed to $1^3\times 128$\\
(8) & Linear, fully connected  & - &  - & - & $1^3 \times 1$ \\ \hline
\end{tabular}
\caption{Architecture of our network. The number of trainable parameters is 563745. The network structure is based on \citet{Mao21}, but we modify it to train with contiguous blocks of data at once by trading the stride with the dilation, such that nonlocal information can be learned and every grid point can be considered in training. 
Note that these modifications result in the apparently asymmetric 
choice of a $2\times2\times2$ kernel in layer 6, but this is mitigated by the additional padding trimmed just before the last fully connected layer.
We explicitly verify that the network produces statistically identical behaviour
when given reversed grids.
We use unpadded convolutional layers for layers 1 through 7. 
The output of layer 7 is originally $5^3$ (per channel), but we manually trim
the grid to $1^3$ (per channel). 
The final layer is a fully connected linear layer that combines the 128 channels
into the prediction of the density field at a single point.
}
\label{tab:architecture}
\end{table*}

We train the model on batches with a fiducial output sub-grid size of $32^3$, corresponding to a padded input 
sub-grid of $70^3$. We explore the sensitivity to 
this choice in Section~\ref{sec:hyper}.
The model parameters are updated every batch. This way, neighbor grid points within the block are considered in the network such that nonlocal (outside the $39^3$ subgrid) information is learned as well. We use Adam optimizer \citep[][]{Kingma14} and start the learning rate at $10^{-4}$ and reduce it by 70\% after every 5 iterations\footnote{One ``iteration'' here means iteration through one simulation. We note the difference from the conventional definition of ``epoch''.} without improvement (we use the \texttt{ReduceLROnPlateau} function of \texttt{Pytorch} \text and set \texttt{factor} to be 0.7 and \texttt{patience} to be 5). 
The learning rate and schedule are chosen by simple experimentation here, and
we defer a detailed exploration of the training schedule and other hyperparameters
to future work.
We set the training to finish when the minimum validation loss does not change for 30 iterations. About half of the cases hit the 48-hours wall time of GPU, but the minimum validation loss has no significant improvement over the last 30 iterations (the minimum validation loss does not change by more than 0.05\%).   
We use the iteration with the lowest validation loss as our model.

Our loss function is the summed MSE loss over all grid points in a batch:
\begin{equation}\label{eqn:loss}
    L=\sum_{i}\bigg[\delta_{\rm CNN}(\boldsymbol{r}_{i})-\delta_{\rm ini}(\boldsymbol{r}_{i})\bigg]^2,
\end{equation}
where $\delta_{\rm ini}$ is our target, the initial condition density. We have tested with an additional regularization term in the loss where we smooth both the CNN output and the target in order to downweight the contribution from small-scale noise. This adjustment, however, led to slightly worse performance in $r(k)$; so we maintain the above simple loss form. We train one simulation (through all subgrids) for one iteration and randomly switch to another simulation for the next iteration. Changing this training schedule, for example, training for 10 iterations before switching simulations, did not improve performance.

\subsection{Simulation and pre-processing of data}
We use the \textsc{Quijote} simulations \citep[][]{Navarro20} in this study. The \textsc{Quijote} simulations are a suite of $N$-body simulations in 1 $h^{-1}$Gpc boxes with various resolutions using a cosmology close to Planck 2018 cosmology \citep[][]{Planck18} as fiducial cosmology and various other cosmologies. We use the 100 high-resolution (1024$^3$ CDM particles) simulations of matter fields with the fiducial cosmology as our fiducial simulations at $z=0$. We test how well our model generalizes both at different redshifts and to different cosmologies from the \textsc{Quijote} Latin Hypercube suite. 

We distribute particles for both the nonlinear and initial condition (at $z=127$) fields on a $512^3$ grid using the triangular-shaped-cloud method \citep[TSC,][]{Hockney88}. We then perform two reconstruction algorithms (detailed in Section~\ref{sect:recon}) on the nonlinear field to obtain reconstructed density field. A smoothing scale of 10$\hMpc$ is used when applying the reconstruction algorithms, unless otherwise noted. We adopt isotropic Gaussian smoothing kernel of convention $S(k)=\exp(-\frac{1}{2}k^2\Sigma_{r}^2)$, where $\Sigma_r$ is the smoothing scale. We use the reconstructed density field for training, together with unsmoothed initial condition as our target. 

The mean of the overdensity is naturally zero, and the reconstructed and initial condition fields are normalized by their standard deviations. This implies that the 
network does not learn the overall normalization of the density field. Physically, this just corresponds to the 
fact that we could choose to reconstruct the linear density field at any redshift, scaling by the growth 
factor. 

We use eight simulations after reconstruction algorithms in training, which corresponds to $8\times512^3$ data points. Using more training data produces minimal improvement. We use two simulations for validation and the remaining 90 simulations for inference \footnote{In a few cases where the noise is low enough or for the analysis using the Latin Hypercube suite where only one snapshot is available, we use only one simulation for inference. These cases are presented in Figures~\ref{fig:diffcos}, \ref{fig:diffz}, \ref{fig:lowdensematter}, \ref{fig:randoms}, and \ref{fig:compare_shallue}.}.

\section{Reconstruction algorithms}\label{sect:recon}
We employ two reconstruction algorithms as inputs to the CNN in this study. \citet[hereafter ES3]{Eisenstein07} estimates the Zel'dovich displacements \citep[][]{Zeldovich70} (the first order solution to the Euler-Poisson system of equations in Lagrangian perturbation theory) of objects and moves them back by these amounts. This method has been used in observations for the last decade and is referred to as the ``standard reconstruction'' in the field. The second method we use is by \citet[hereafter HE18]{Hada18}. This method does not move particles but reconstructs the density field directly instead. It does this iteratively using an annealing smoothing scheme, such that the smoothing scale starts at a fixed, relatively large, value but reduces with iterations to an ``effective'' smoothing scale and stays at the scale afterwards. This allows it to 
(approximately) solve for the displacement as a function of Lagrangian coordinates (as opposed
to Eulerian coordinates in ES3).
Furthermore, it adopts the second order solution of the Euler-Poisson system of equations. A detailed comparison of the two algorithms is in an upcoming paper \citep[][]{reconpaper}. 

A choice in reconstruction is whether to reconstruct the linear density field in real or redshift space \citep[e.g.][]{Seo10,Seo16b,Chen19}. We adopt the isotropic convention here, in that we choose to reconstruct real space density field, i.e. we remove redshift space distortions in the reconstruction process.

\section{Results and Robustness}\label{sect:results}
We present our results on the performance of the neural network applied to the real and redshift space density field at $z=0$ (Section~\ref{sect:defaultcos}). Section~\ref{sect:diffcos} presents our tests for the robustness of our results to cosmology and redshift. Section~\ref{sect:shotnoise} explores the degradation of this reconstruction with increasing shot noise, describing a simple model to explain the observed trend.

We measure the efficacy of reconstruction using propagator and cross correlation coefficient defined as 
\begin{equation}
G(k)=\frac{\left<\delta^{*}(k)\delta_{\rm ini}(k)\right>}{\left<\delta^2_{\rm ini}(k)\right>} \end{equation}
and
\begin{equation}
r(k)=\frac{\left<\delta^{*}(k)\delta_{\rm ini}(k)\right>}{\sqrt{\left<\delta^2(k)\right>\left<\delta^2_{\rm ini}(k)\right>}},
\end{equation}
respectively. In the above, $\delta(k)$ and $\delta_{\rm ini}(k)$ are the reconstructed and the initial densities in Fourier space, respectively. Both $G(k)$ and $r(k)$ represents correlation with the initial density field, but $G(k)$ retains amplitude information of the field while $r(k)$ is indicative of purely the phase. The propagator is widely used in BAO analyses, because it relates to the damping and the BAO feature due to nonlinear evolution and is used to assess the effectiveness of reconstruction. We sometimes refer to both propagator and the cross-correlation coefficient as ``cross-correlations''. 
The goal of reconstruction is to bring these cross-correlations as close to 1 as possible.

We also compare the power spectra of the reconstructed field to linear theory. We measure the power spectrum multipoles defined as
\begin{equation}
    P_{l}(k)=\frac{2l+1}{2}\int_{-1}^{1}P(k,\mu)L_{l}(\mu)\de \mu,
\end{equation}
where $L_{l}$ is the Legendre polynomial of order $l$. In real space, only the $l=0$ multipole (i.e. monopole) is nonzero. We use the quadrupole ($l=2$) to assess how well reconstruction has removed redshift space distortions.

\subsection{Fiducial results}\label{sect:defaultcos} 
\begin{figure}
    \centering
    \includegraphics[width=\columnwidth]{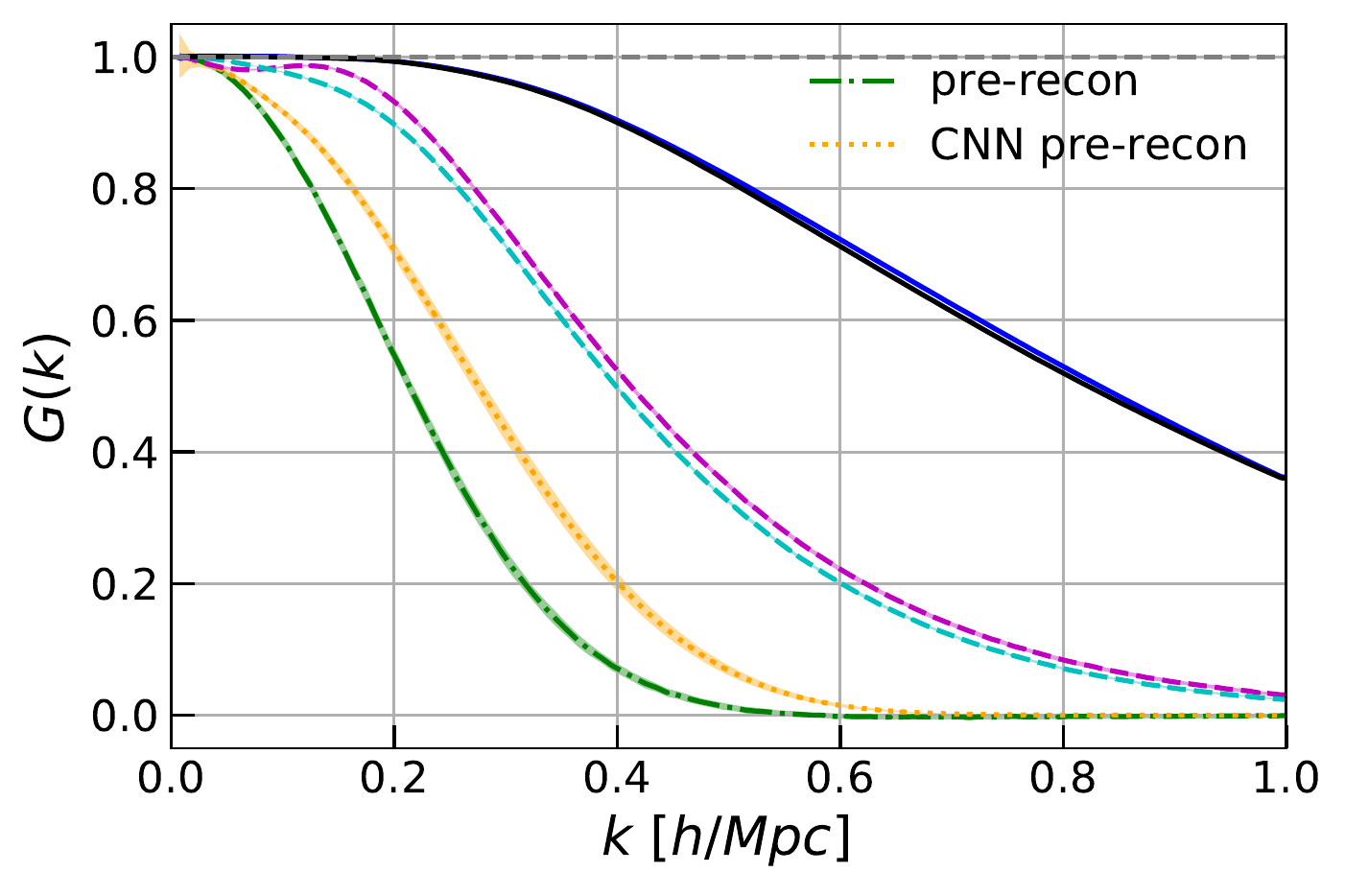}
    \includegraphics[width=\columnwidth]{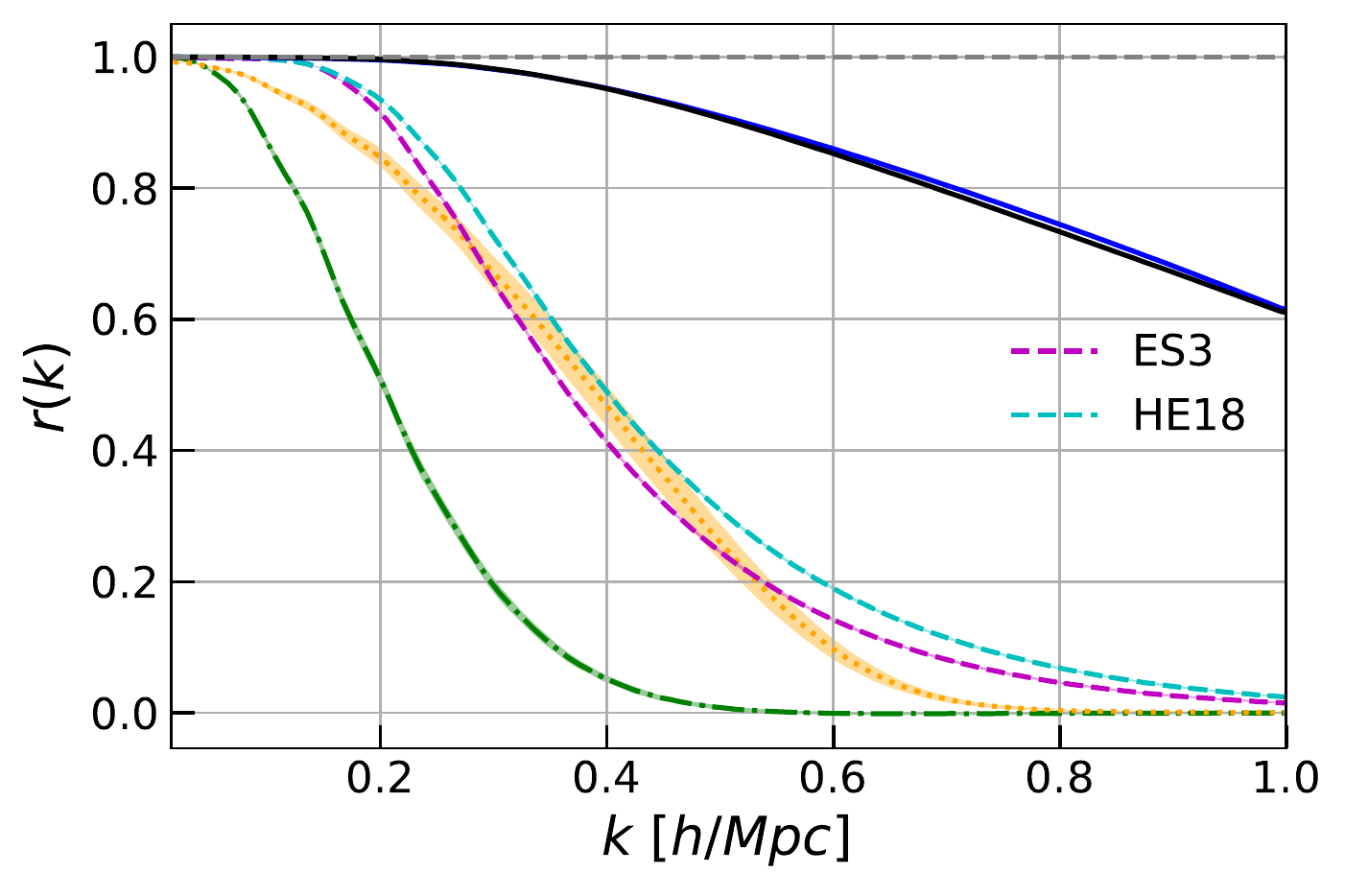}
    \includegraphics[width=\columnwidth]{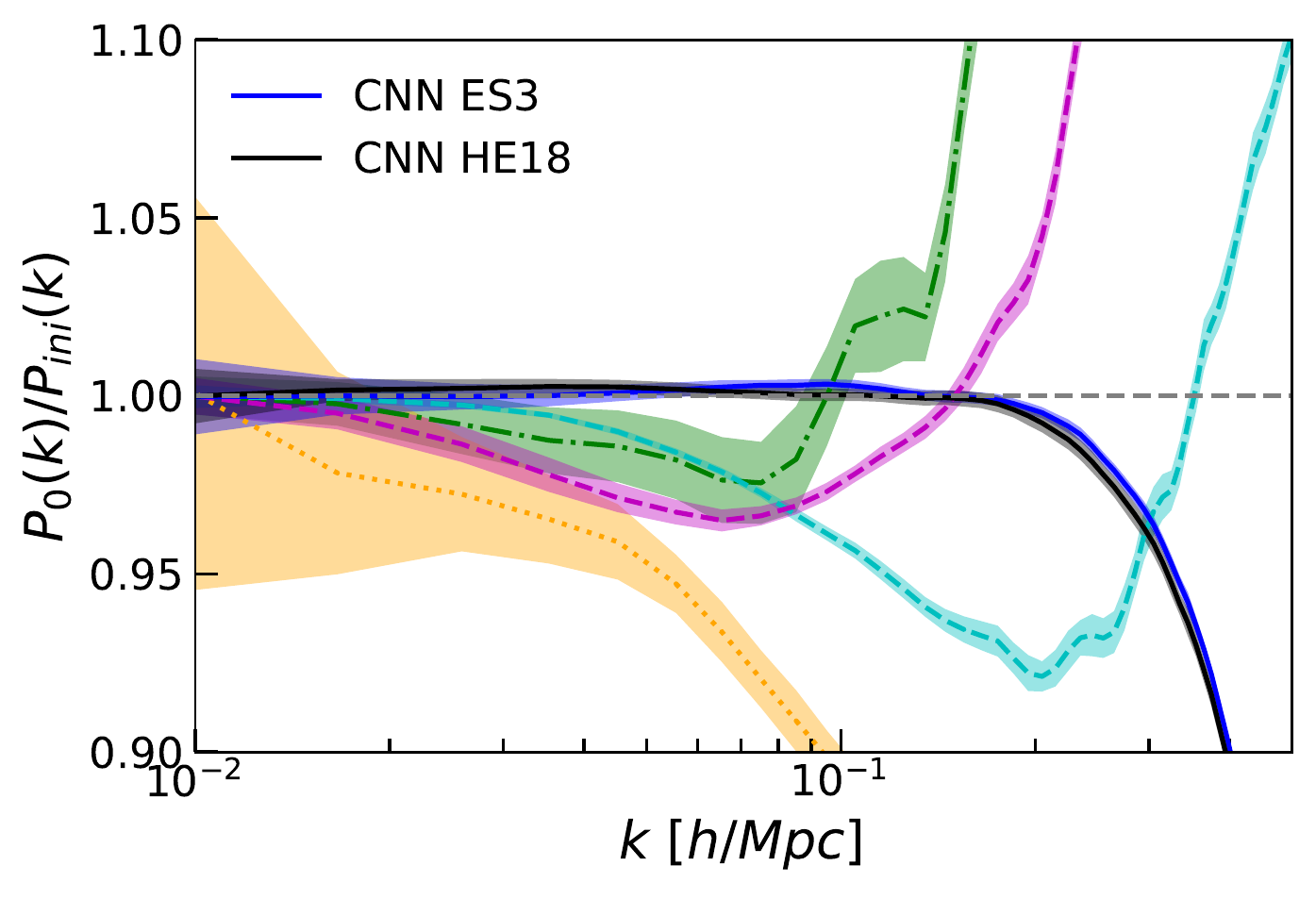}
    \caption{The performance of reconstruction measured by the propagator (upper), 
    cross-correlation coefficient (middle), and the monopole power spectrum (bottom) at $z=0$ in real space.
    We focus on our models - CNNs trained with reconstructed fields by ES3 reconstruction (blue solid) and by HE18 reconstruction (black solid). 
    Also presented for comparison are the input ES3 reconstruction (magenta dashed) and HE18 reconstruction (cyan dashed) fields, 
    a CNN trained with $z=0$ nonlinear (i.e. unreconstructed) density field (yellow dotted) and $z=0$ nonlinear field itself (green dash-dotted). 
    Each line is an average of 90 simulations that were not used in any of the training steps,
    with the shaded region showing the $1-\sigma$ region.
    The $G(k)$ and power spectrum curves for the CNNs were rescaled to 1 on the largest scales ($k=0.01h\Mpc^{-1}$), but this correction was less
    than 0.5\%, except for the network run on unreconstructed data (see the text for further discussion).
    The improvements here are clear, with the CNN reconstructed density fields closely following the initial density field 
    to much higher $k$ than previously. Note that the power spectrum is plotted on a different scale to highlight the agreement 
    on scales $k < 0.2 h\Mpc^{-1}$; the behaviour at higher $k$ is similar to (and consistent with) $r(k)$ and $G(k)$.
    }
    \label{fig:defaultcos}
\end{figure}

Figure~\ref{fig:defaultcos} shows the performance of the CNN-augmented reconstruction, compared to the reconstruction algorithms alone, at $z=0$ in real space. We also show the output of a CNN using the pre-reconstruction density field as input 
(as in \citet{Mao21}) 
\footnote{For $G(k)$ and the power spectrum, the amplitude is rescaled to 1 at the largest scale, $k=0.01h\Mpc^{-1}$. The rescaling factor in $G(k)$ for all the reconstruction cases and pre-reconstruction are between 0.997 and 0.998 and is 1.016 for CNN pre-reconstruction. The range of rescaling factor for power spectrum is between 0.9945 -- 1.0004, except for CNN with pre-reconstruction, which is rescaled by a factor of 1.064.}. 
In all these metrices, a CNN trained with reconstructed density fields achieves significantly better results than reconstruction algorithms alone or a CNN trained with pre-reconstruction fields.
For the cross-correlation, $r(k)\geq 0.9$ is maintained up to $k\sim0.5 h\Mpc^{-1}$ for CNN with ES3 as well as HE18, a factor of $\sim$2 improvement from those of reconstruction algorithms alone. Similar improvements apply to $G(k)$, with $G(k)\geq 0.9$ preserved up to $k\sim0.4 h\Mpc^{-1}$.
Applying a CNN to the pre-reconstruction density fields does generally worse than reconstruction algorithms alone.
The network reconstructs the power spectra to better than 95\% out to $k\sim 0.3 h\Mpc^{-1}$ and 90\% out of $k\sim 0.4 h\Mpc^{-1}$.

\begin{figure}
    \centering
    \includegraphics[width=\columnwidth]{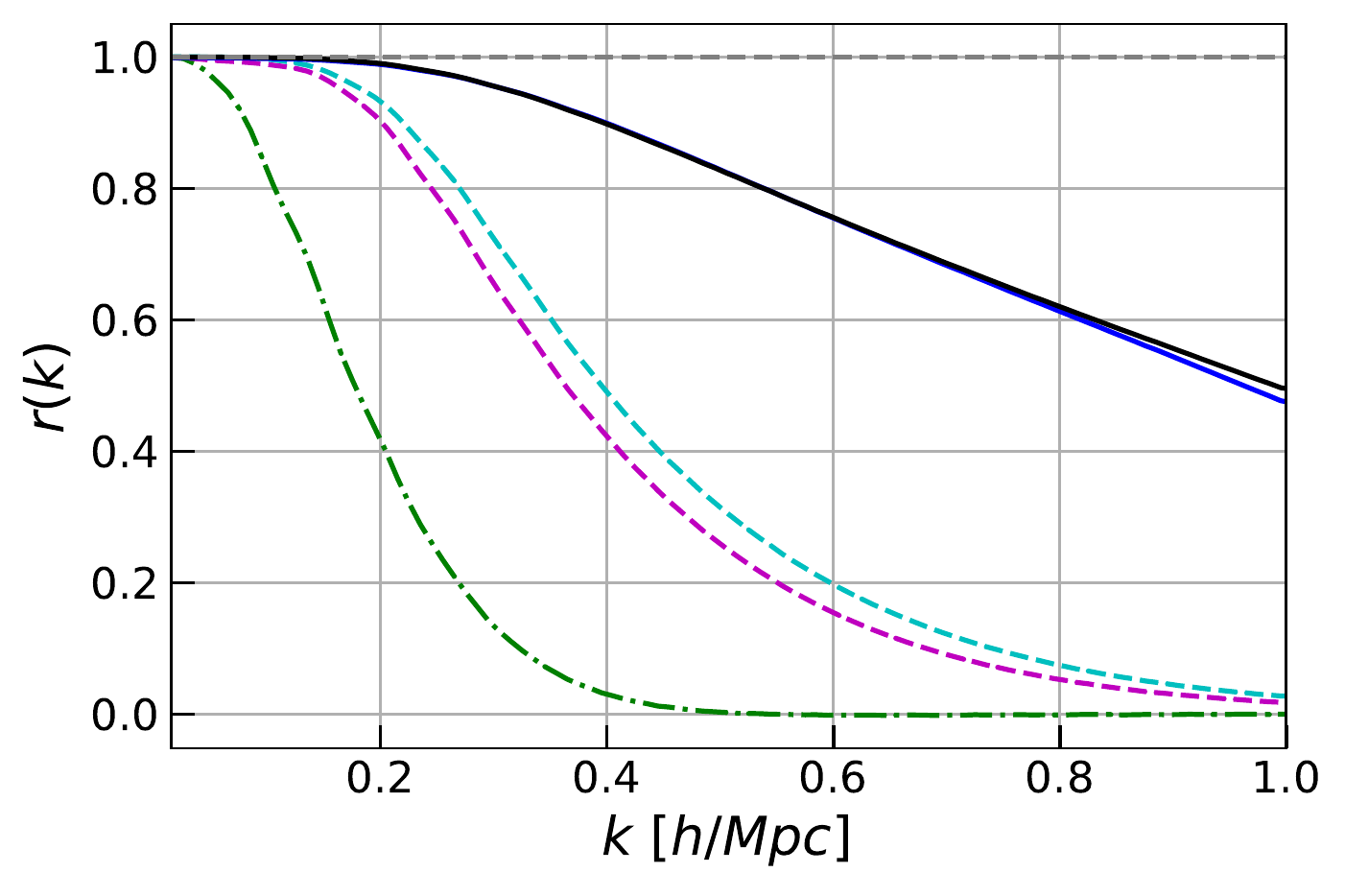}
    \includegraphics[width=\columnwidth]{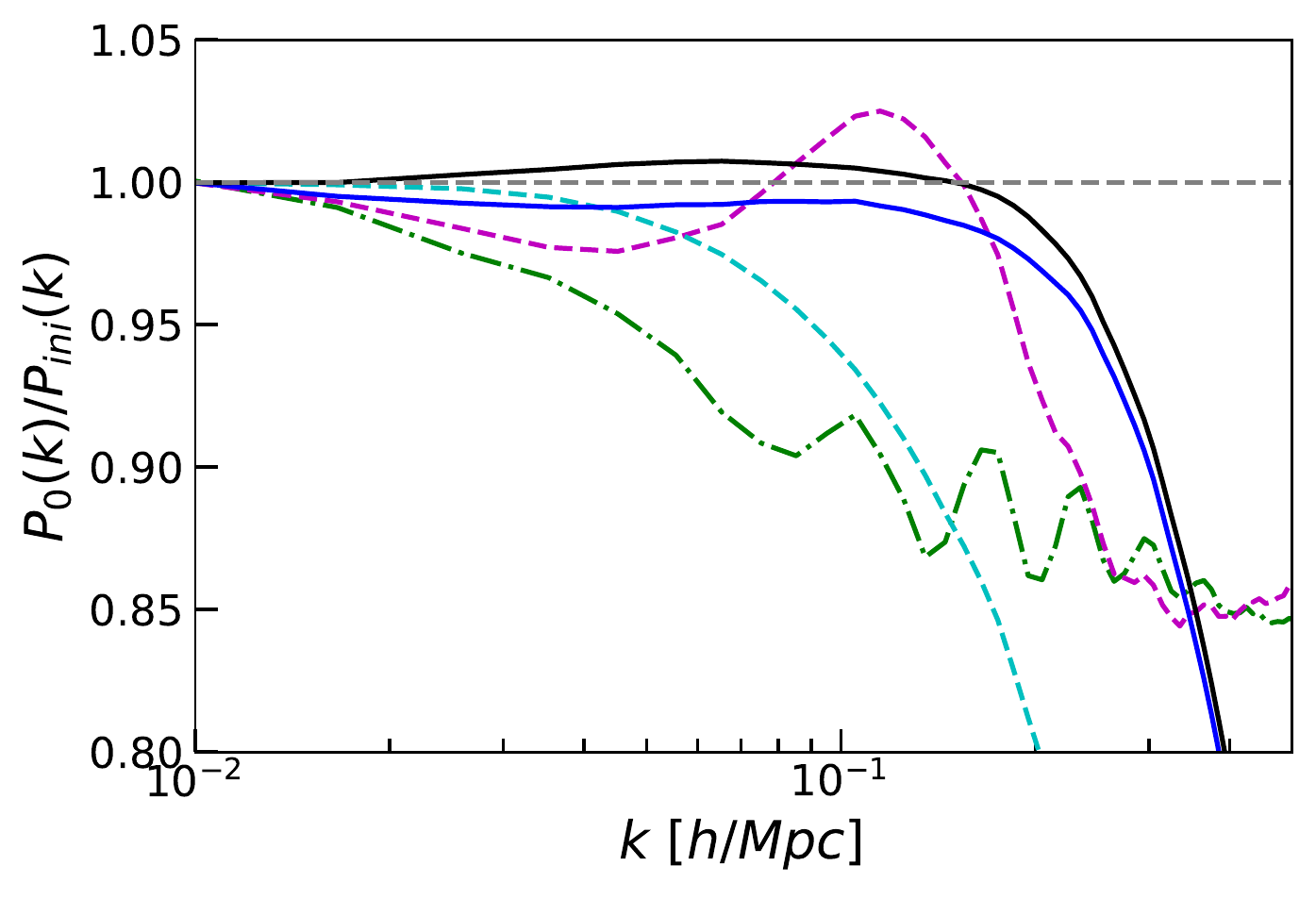}
    \includegraphics[width=\columnwidth]{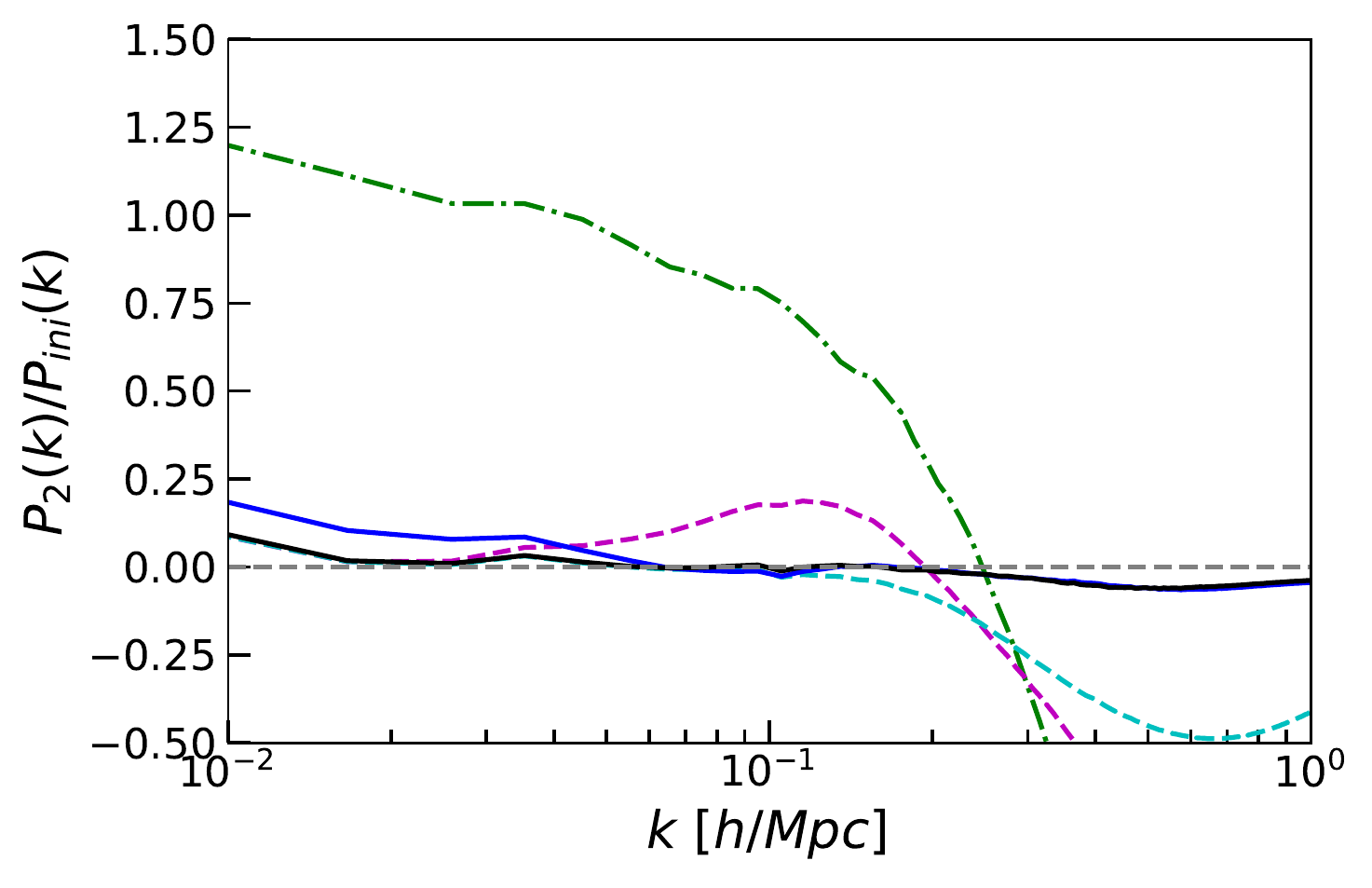}
    \caption{The reconstruction performance in redshift space at $z=0$ measured through the cross-correlation coefficient (top) 
    and monopole (middle) and quadrupole (bottom) power spectrum of redshift space model. The line styles follow the 
    convention in Figure~\ref{fig:defaultcos}. As in the case of real-space fields, the neural network significantly 
    boosts the performance of the reconstruction algorithm. We also note that it improves on the removal of redshift-space 
    distortions, with the HE18 input doing slightly better.
    }
    \label{fig:zspace}
\end{figure}

We train a separate model for redshift space at $z=0$, using redshift space data and the same CNN architecture. Figure~\ref{fig:zspace} shows the performance of CNN compared to reconstruction algorithms, as in Figure~\ref{fig:defaultcos}\footnote{In monopole power spectrum, every line is rescaled to match 1 on large scale, $k=0.01 h\Mpc^{-1}$. The rescaling factor is between 0.993 and 1.005. Quadrupole power spectra are not rescaled.}. As in real space, the CNN trained with reconstructed data performs much better than reconstruction algorithms alone. 
The HE18 version restores the monopole power spectrum slightly better than the ES3 version, even if the HE18 power spectrum is far off. The two CNNs are comparable for quadrupole power spectrum. Both are very close to zero on large $k$, suggesting that CNN can effectively remove redshift space distortions on small scales. However, on large scales, CNN with ES3 reconstruction produces spurious power, i.e., output worse than input. This finding echos the observation in \citet{Shallue22}. 
The HE18-CNN, however, performs as well as HE18 on large scales. 
Finally, the CNN propagators do not produce the bump-like feature near $k=0.1 h\Mpc^{-1}$ seen in  
standard reconstruction algorithms (see e.g. Figure~10 in \citet[][]{Seo22} and Figure~1 in \citet[][]{Seo16b}) for certain choices
of the smoothing scale.
These results suggests that both CNNs can successfully reconstruct the real space density field on most scales, with 
the ES3-CNN slightly compromising the large scales.

\subsection{Different cosmology and redshift}\label{sect:diffcos}
To examine the robustness of our model, we test it with simulations of different cosmologies and a different redshift. We use eleven different cosmologies in \textsc{Quijote}'s high-resolution Latin Hypercube simulations, where $\Omega_{\rm m}$ $\in$ [0.1755, 0.4985], $\Omega_{\rm b}$ $\in$ [0.6233, 0.9865], $h$ $\in$ [0.5265, 0.8883], $n_s$ $\in$ [0.8307, 1.1375], and $\sigma_8$ $\in$ [0.6233, 0.9865]. For reference, the fiducial cosmology has ($\Omega_{\rm m}$, $\Omega_{\rm b}$, $h$, $n_s$, $\sigma_8$)=(0.3175, 0.0490, 0.6711, 0.9624, 0.8340). We also apply our model to simulations at $z=1$, which represents a more drastic change of cosmology, where $\Omega_{\rm m}$ increases from 0.3175 to 0.7882.

\begin{figure}
    \centering
    \includegraphics[width=\columnwidth]{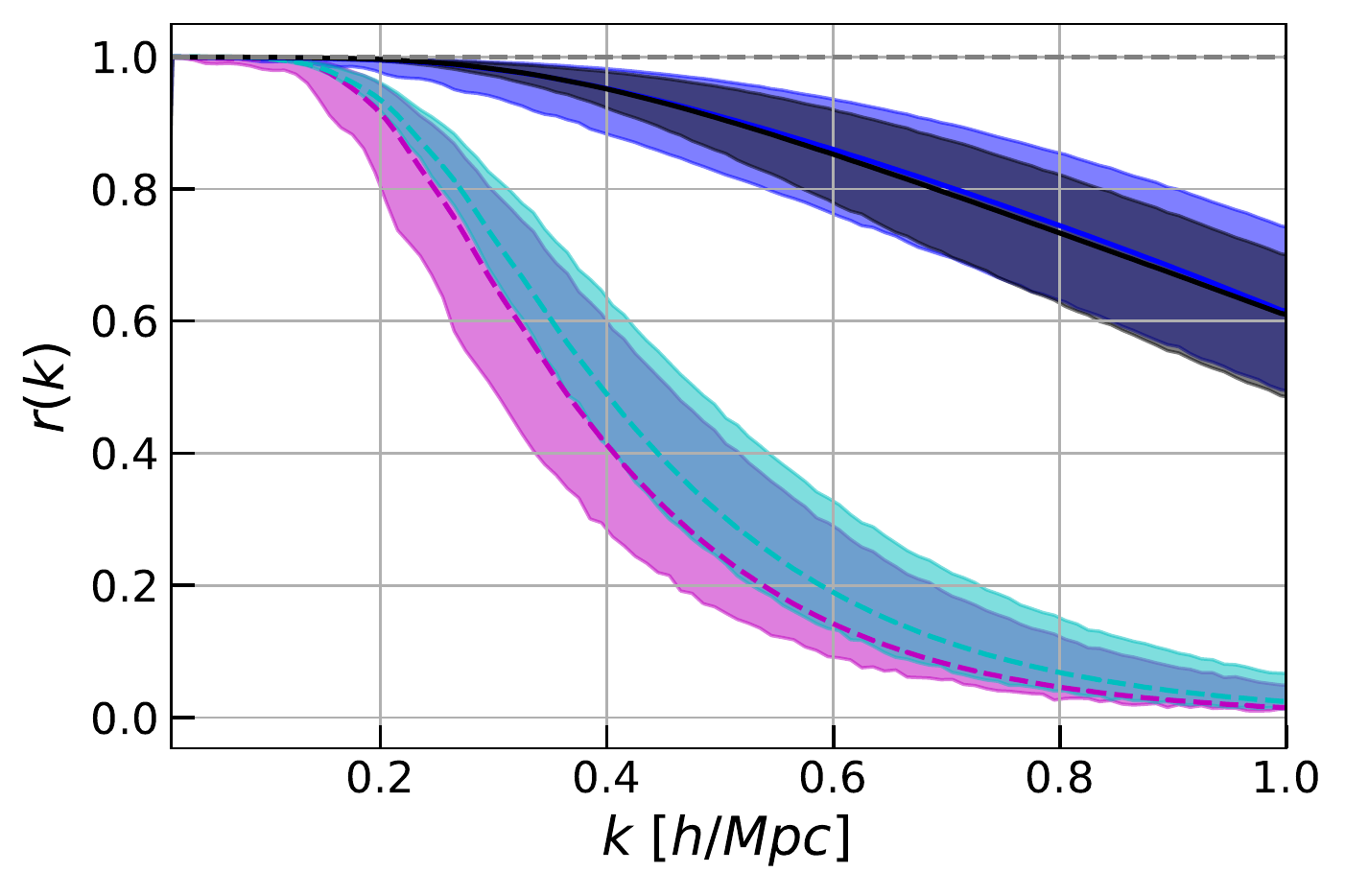}
    \caption{The bands show the spread in cross-correlation coefficients $r(k)$ for a wide range of
    cosmologies at $z=0$. In all cases, the CNN models were trained on simulations from the fiducial cosmology, 
    so this figure demonstrates how the model generalizes across cosmology. The cyan and black bands show $r(k)$
    for the standard HE18 and CNN-augmented HE18 reconstructions respectively, while the magenta and blue bands 
    are the analogous bands for the ES3 reconstruction. The dashed and solid lines show the performance on the
    fiducial cosmology (that the CNN models were trained on) for reference.
    } 
    \label{fig:diffcos}
\end{figure}

We find that the model trained with one fiducial cosmology can be applied to datasets with different cosmologies and still improve upon the input reconstruction algorithms. 
As shown in Figure~\ref{fig:diffcos}, for these variations in cosmology, the two input reconstruction algorithms produce wide bands in $r(k)$, with that of ES3 wider than that of HE18. The variation 
after processing with the CNN is also slightly wider in ES3 reconstruction than in HE18. Note that the edge of the band for CNN does not necessarily correspond to the edge of the band for the algorithm.

\begin{figure}
    \centering
    \includegraphics[width=\columnwidth]{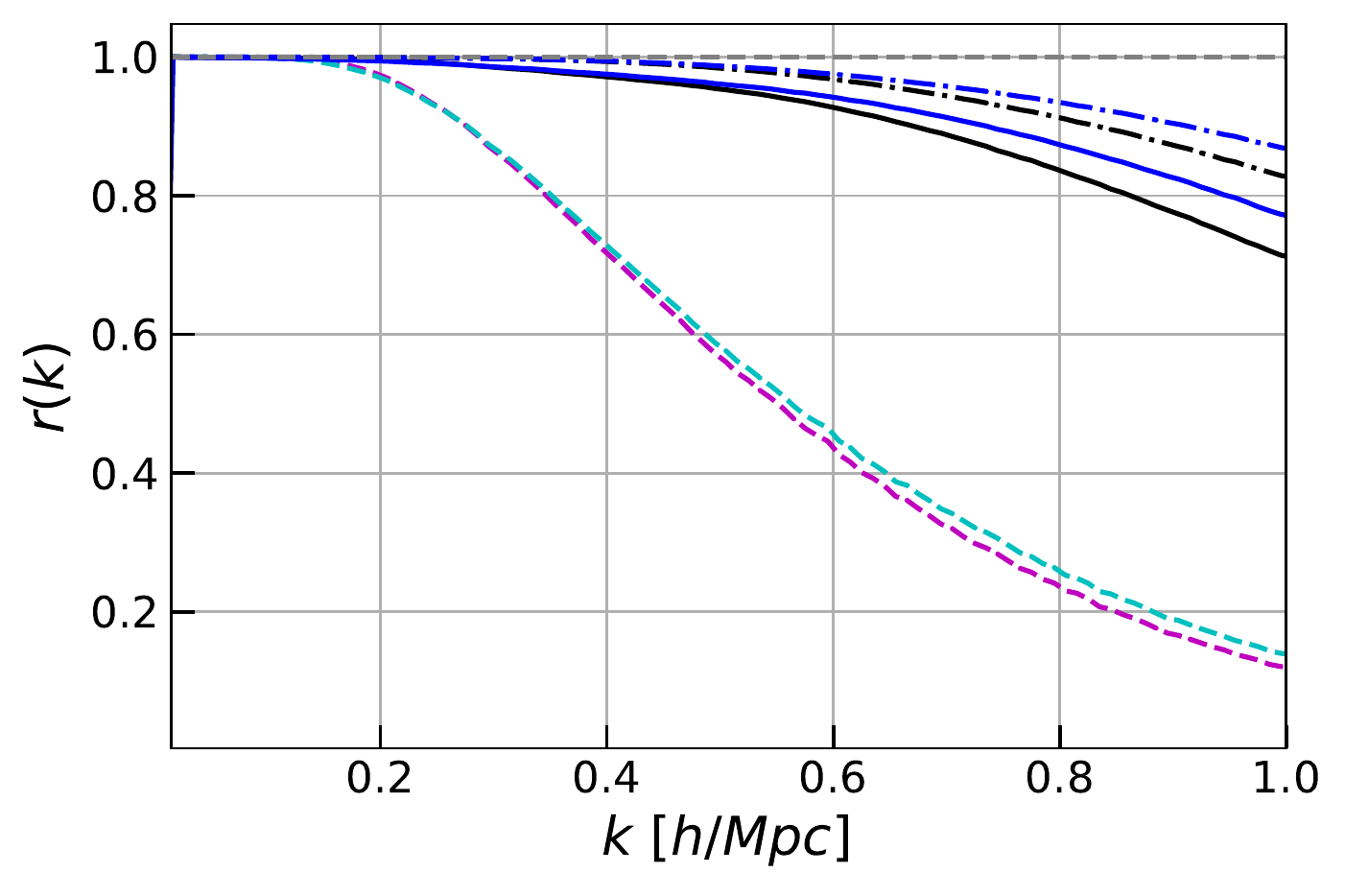}
    \caption{The solid lines show the performance of reconstruction when a model trained on a $z=0$ density 
    field is applied to a $z=1$ density field (with black and blue corresponding to the HE18 and ES3 models). 
    The dot-dashed lines show the improvement when the models are trained on $z=1$ density fields, while the 
    dashed lines (cyan and magenta corresponding to HE18 and ES3, respectively) show the input reconstruction 
    performance. We find that the neural network learns redshift-independent features that allow it to reconstruct the density 
    field significantly better than the standard methods, although training with data more closely matched 
    to the input redshift still provides some further improvements.
    } 
    \label{fig:diffz}
\end{figure}

Applying our model to a different redshift, we find that the network is robust, even though a model trained with proper redshift will do better. The prediction for $z=1$ data is noticeably worse compared to a model trained with $z=1$ simulations, shown in Figure~\ref{fig:diffz}. In this case, ES3-CNN performs better than HE18-CNN. Interestingly, this observation is mirrored in 
some of the cosmologies 
in the Latin Hypercube test above where ES3 achieves better application results with lower $\sigma_8$ and high $\Omega_{\rm m}$.

These results show that the neural network is learning features that are generic and not tied to the cosmology used for training. This also suggests that the network will be robust even if the training does not match the observed data perfectly (as one might expect). An interesting open challenge could be to determine what information/features the network is using to reconstruct and if these can be explained.

\subsection{Shot noise}\label{sect:shotnoise}

The above results were all for the matter density field with effectively no shot noise. In this section, we consider how reconstruction degrades in the presence of shot noise. We consider number densities between $10^{-4}$ $h^3\Mpc^{-3}$ and $10^{-1}$ $h^3\Mpc^{-3}$ by randomly subsampling the matter field (The full field has a number density of $\sim$1 $h^3\Mpc^{-3}$). 
An interesting choice here is whether to apply the model trained on the (effectively)
noiseless field, or to retrain the model on the noisy data. We test both 
variants and find that applying the model trained on the noise-free data performs
slightly worse than the retrained model. Our results below therefore focus on 
the cases where the CNN was retrained.

We characterize the performance of the reconstruction by quoting the $k$-value at which either the cross-correlation coefficient $r(k)$ or the propagator $G(k)$ drops to 0.7. The particular choice of level is arbitrary. The further in $k$ this value goes, the more effective the reconstruction is.

\begin{figure}
    \centering
    \includegraphics[width=\columnwidth]{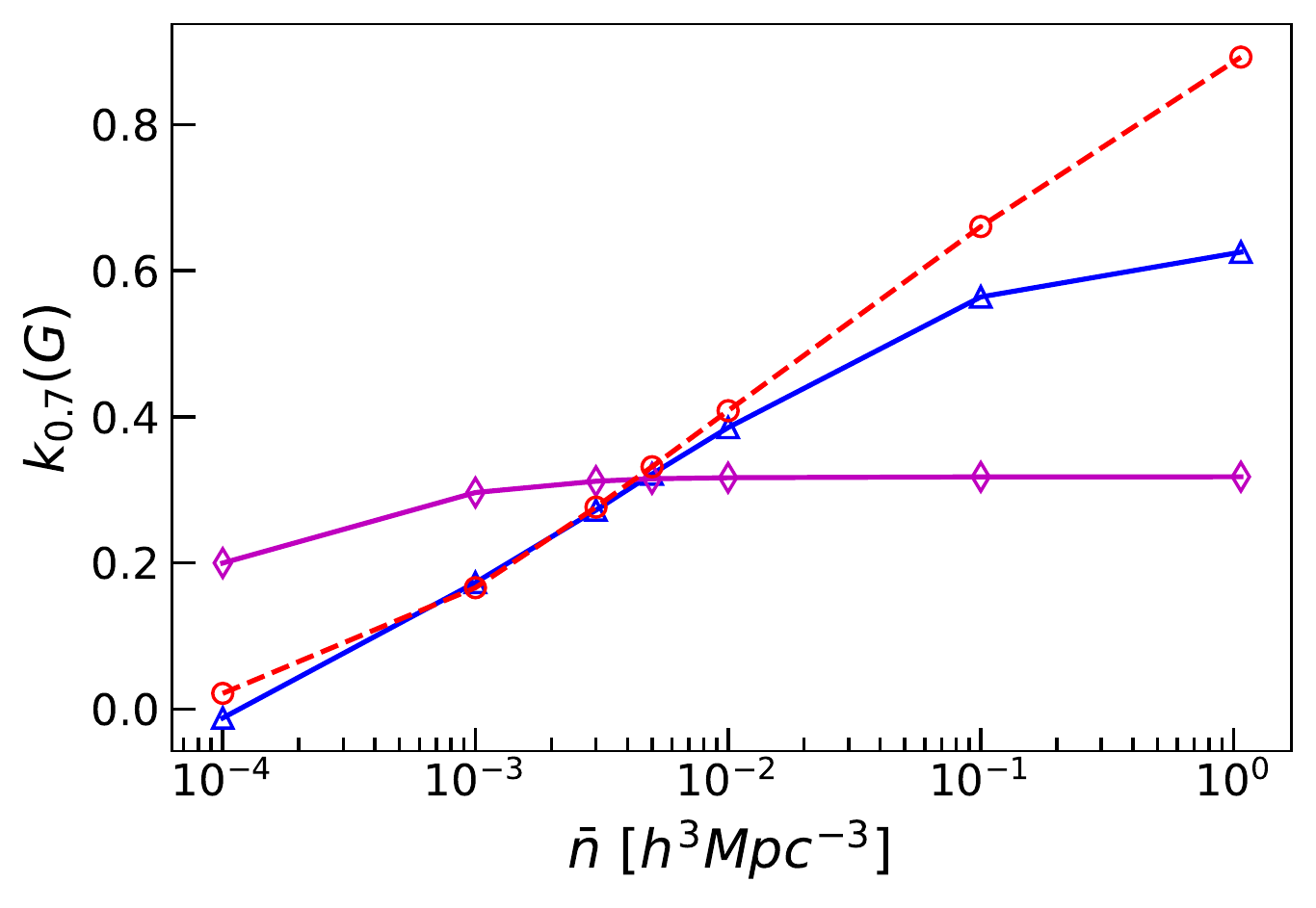}
    \includegraphics[width=\columnwidth]{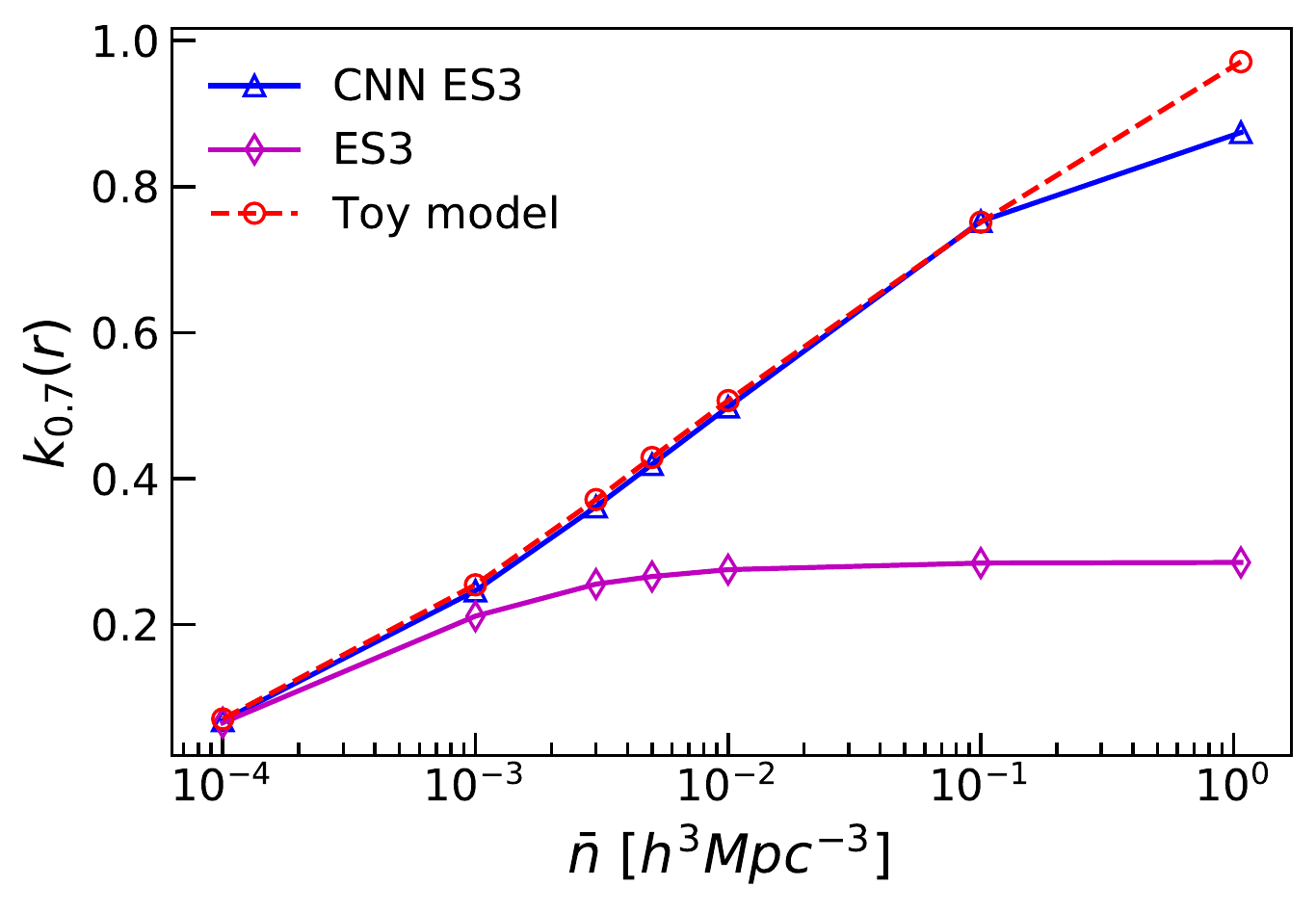}
    \caption{The $k$ value where $G(k)$ (upper) and $r(k)$ (lower) drop to 0.7 as a function of number density, for ES3 (magenta diamond) and ES3-CNN (blue triangle) reconstruction. 
	We see that the input reconstruction appears to saturate at about $\bar{n}=10^{-3} h^3\Mpc^{-3}$
	for $k(G)$ and about $10^{-2}h^3\Mpc^{-3}$ for $k(r)$, while the ES3-CNN starts to slow down the improvement around 
    $10^{-1}h^3\Mpc^{-3}$. Also shown is a toy model (see text for details) which captures the 
	expected degradation of the CNN's performance with increasing shot noise.  
For simplicity, we only show ES3 results here, but HE18 has similar patterns. 
    }
    \label{fig:lowdensematter}
\end{figure}

Figure~\ref{fig:lowdensematter} shows the $k$ values where $G(k)$ and $r(k)$ equal 0.7. We only show the ES3 versions for simplicity, but HE18 results are similar. The dependence on the number density of the field for the input reconstruction algorithm saturates above a number density of $10^{-3}$ $h^3\Mpc^{-3}$ for $G(k)$ and $10^{-2}$ $h^3\Mpc^{-3}$ for $r(k)$. The dependence for $r(k)$ is also fairly mild between $10^{-3}$ $h^3\Mpc^{-3}$ and $10^{-2}$ $h^3\Mpc^{-3}$. This confirms the finding by \cite{white10} using Lagrangian perturbation theory. 
By contrast,
the CNN-reconstructed field continues to improve with increasing number density, appearing to slow down around $\Bar{n}=10^{-1}h^3\Mpc^{-3}$. This is not surprising since the neural network is 
presumably learning features smaller than where the Zel'dovich approximation would be valid.

To quantitatively understand this behavior, we construct a simple model for the CNN reconstruction model here.
We assume that the input to the CNN is $\delta_{\rm in}=\delta_{\rm L}+\epsilon$, where $\delta_{\rm L}$ is the linear density field, and $\epsilon$ is the shot noise assumed to be uncorrelated with the density field. The target field is the linear density field, $\delta_{\rm L}$. 
We further assume that the estimated field, $\hat{\delta}({\boldsymbol{k})}$, is a linear function of $\delta_{\rm in}(\boldsymbol{k})$: 
\begin{equation}\label{eqn:delta_hat}
\hat{\delta}({\boldsymbol{k})}=W(k)(\delta_{\rm L}(\boldsymbol{k})+\epsilon). 
\end{equation}
Note that this is a crude approximation, ignoring both the locality and nonlinearities of the the CNN. We solve for $W(k)$ by minimizing the loss function (see Appendix~\ref{appx:Wiener} for details): 
\begin{equation}
W(k)=\frac{P_{\rm L}(k)}{P_{\rm L}(k)+\frac{1}{\Bar{n}}},
\end{equation}
where $\frac{1}{\Bar{n}}$ is the shot noise power. The power spectrum $P_{\rm L}(k)$ here is the linear power spectrum modulated by the effect of the grid, which we approximate for simplicity by $W_{\rm grid}(k)=\exp(-k^2R_{\rm g}^2)$, where $R_{\rm g}=1.95\hMpc$, our grid spacing. 

With this setup, we compute $G(k)$ and $r(k)$:
\begin{equation}\label{eqn:theory_gkrk}
    \begin{aligned}
    G(k)&=\frac{P_{\rm L}(k)}{P_{\rm L}(k)+\frac{1}{\Bar{n}}},\\
    r(k)&=\sqrt{\frac{P_{\rm L}(k)}{P_{\rm L}(k)+\frac{1}{\Bar{n}}}}.
    \end{aligned}
\end{equation}
In this simple model, we have $G(k)=r(k)^2$. Since $|r(k)|\leq1$, this relation indicates that $G(k)<r(k)$ in general, which is what we observe.

As we see from the figures, this simple model faithfully captures the behavior of the CNN at low number densities. The deviation at higher number densities is also not surprising. The true small-scale density field input to the CNN does not carry the linear field information that our toy model assumes.

These results highlight a major requirement of such CNN-based reconstruction approaches -- the need for high number density samples. Indeed, based on our simple toy model, we would expect similar behavior for any improved reconstruction scheme. Of course, this would also be modified by biased tracers, which could boost the signal relative to the shot noise. We defer those studies to later work.

\section{Discussion}~\label{sect:discussion}

The above has focused on the basic performance of the CNN-based reconstruction. We now turn to a discussion of the robustness of our approach as well as a comparison to previous work.

\subsection{Sensitivity to input reconstruction}

The ES3 and HE18 reconstruction algorithms coupled with a CNN show comparable levels of improvement in $G(k)$, $r(k)$ and power spectrum. However, HE18-CNN performs better in redshift space and it is slightly more robust. A significant benefit of HE18-CNN is its removing of the RSD. In the redshift space quadrupole, the HE18-CNN does not introduce more power on large scales. 
One reason here might be that HE18 does attempt to find the displacement in Lagrangian 
(as opposed to Eulerian) space.
We reserve more investigation in this direction to future work.

\begin{figure}
    \centering
    \includegraphics[width=\columnwidth]{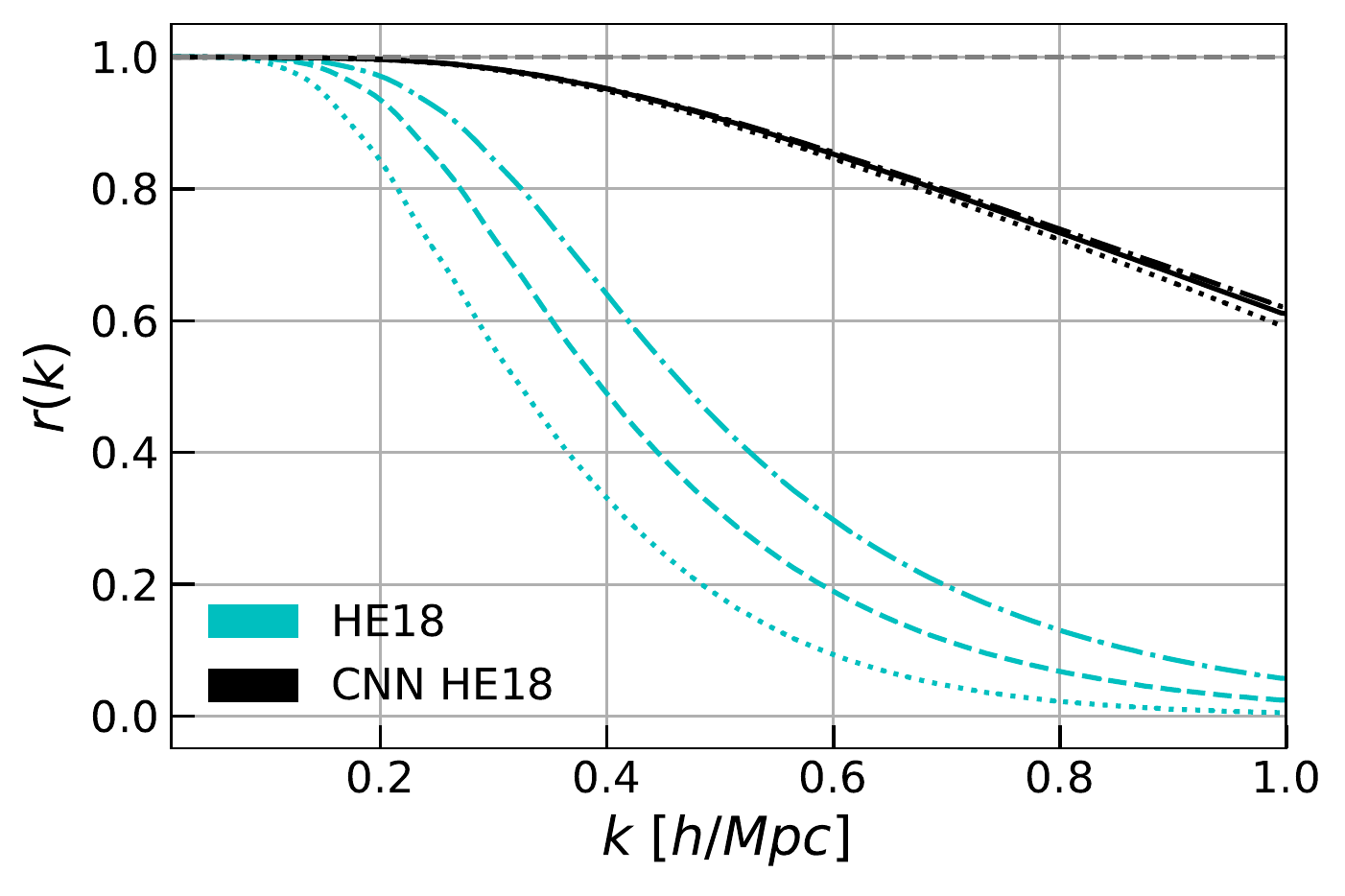}
    \includegraphics[width=\columnwidth]{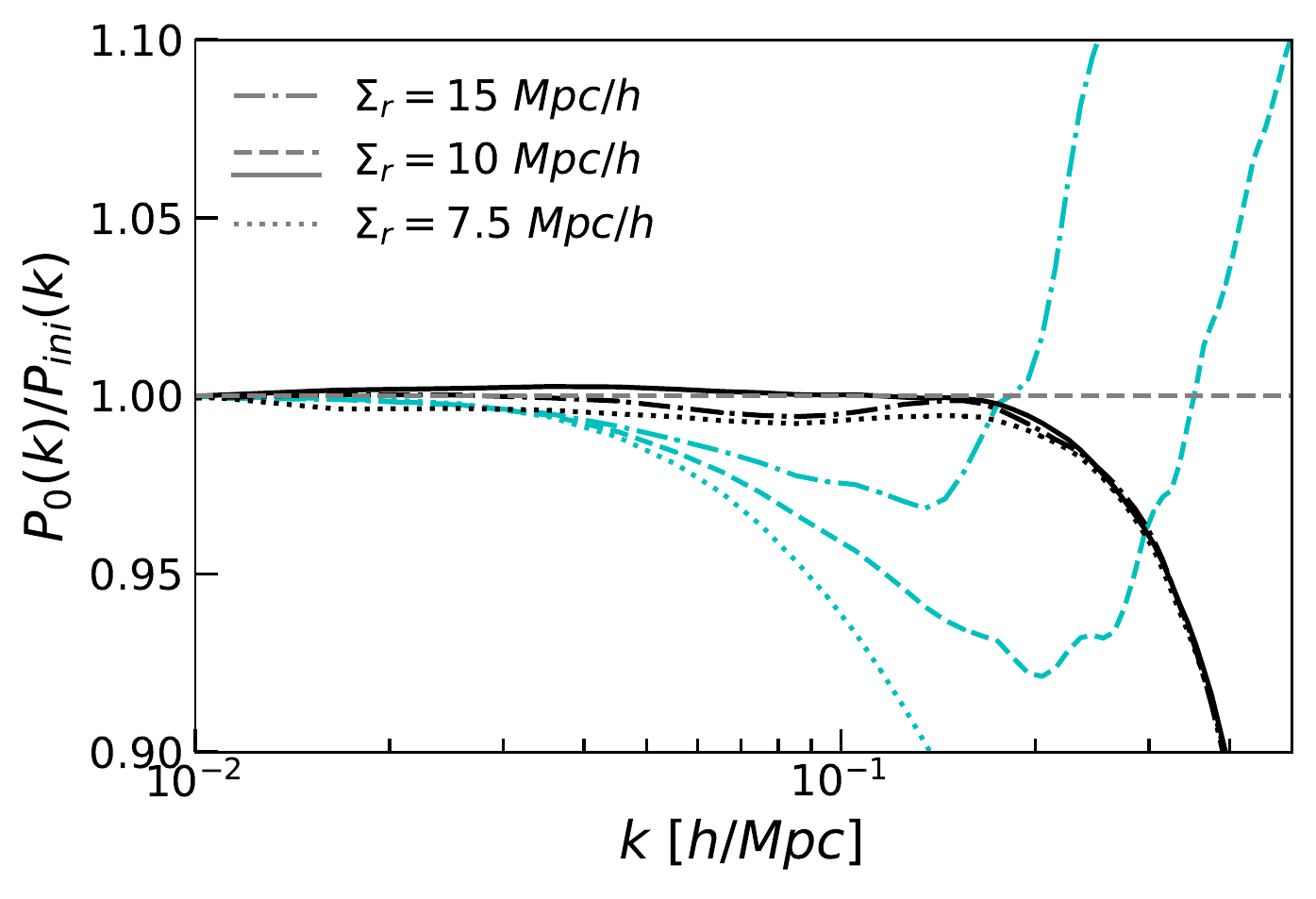}

    \caption{  
    A demonstration that the CNN-augmented reconstruction can significantly reduce the variance that arises from 
    different choices of input reconstruction parameters. We focus here on the input smoothing scale, which has 
    the largest impact on the reconstruction. The cross-correlation coefficients (upper panel) and power spectra (lower panel) for 
    smoothing scales of 7.5 (dotted), 10 (dashed (pre-CNN) and solid (post-CNN) to match Figure~\ref{fig:defaultcos} convention) and 15$\hMpc$ (dash-dotted) show much less scatter after the CNN-augmented reconstruction 
    (black lines) than before (cyan). For simplicity, we show only the HE18 variant, but the ES3 version performs similarly. 
    }
    \label{fig:diffsmoothing}
\end{figure}

In the above tests, we use a smoothing scale of 10$\hMpc$ for both reconstruction algorithms. We now show the effect of changing the smoothing scale
\footnote{For all the tests described here, we retrain the neural network with the new parameters.}.
Figure~\ref{fig:diffsmoothing} shows $r(k)$ and monopole power spectrum of the HE18 algorithm and its corresponding CNN outputs with smoothing scales in the range of 7.5 -- 15$\hMpc$ \footnote{For the power spectrum, the magnitude is rescaled to 1 on large scale, with the rescaling factor between 0.9920 and 1.0045.}. For simplicity, we only show HE18 
in real space here, but other variants behave similarly. 
The different smoothing scales as well as the different algorithms result in a wide range in $r(k)$ and power spectrum for the input, but the CNN largely removes the dependence of the details in the input reconstruction.

\begin{figure*}
    \centering
    \includegraphics[width=\columnwidth]{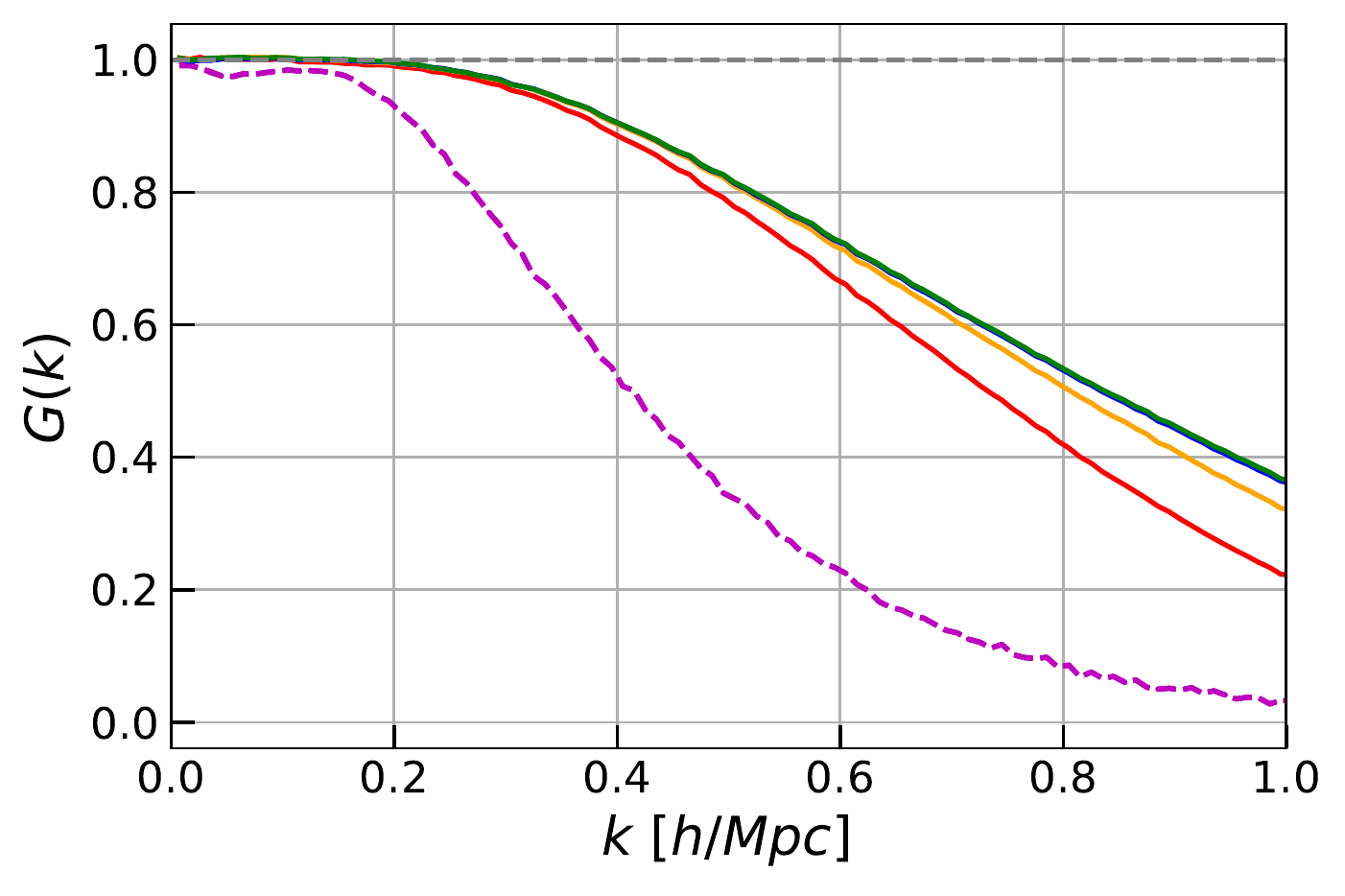}
    \includegraphics[width=\columnwidth]{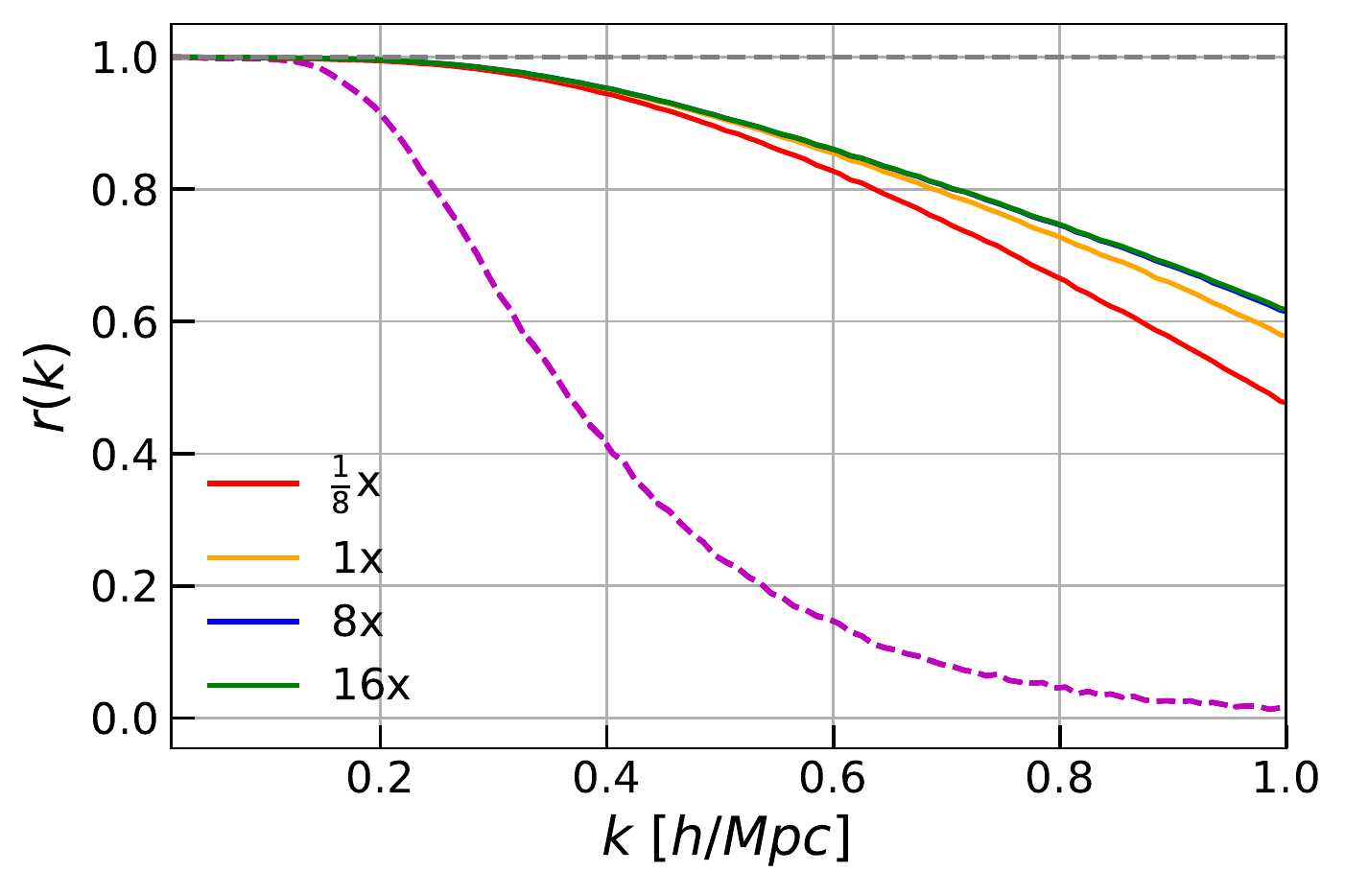}

    \caption{Performance of ES3-CNN in $G(k)$ (left) and $r(k)$ (right) when
    the number of randoms used in the ES3 algorithm varies. 
	The solid lines show the CNN performance for different number densities (described
	in the legend), while the dashed lines are the corresponding ES3 results (all of which
    completely overlap in the figure). 
	These improvements are consistent with the picture that the CNN is 
    reconstructing based on features at small scales and is therefore more sensitive to 
	noise on those scales.
    }
    \label{fig:randoms}
\end{figure*}

A parameter in the ES3 and ES3-CNN tests is the number of random particles employed in reconstruction. Increasing the number of randoms is effectively reducing the shot noise in the field. 
We show the effects of number of randoms in Figure~\ref{fig:randoms}, where we present $G(k)$ and $r(k)$ from training with ES3 reconstructed fields in real space when the number of randoms used in reconstruction varies. We find that when the number of randoms is decreased by a factor of 8 from $1024^3$ (the number of particles in the simulation) to $512^3$, the cross-correlation of the CNN significantly drops. The benefit from increasing the randoms from $1024^3$ quickly saturates. The differences only appear in CNN, with the results from input ES3 reconstruction not sensitive
to this. We use $1024^3\times 8$ randoms throughout this work. This behaviour can be understood in the 
context of the discussion in Section~\ref{sect:shotnoise} and Figure~\ref{fig:lowdensematter}. 
A decrease in the number of randoms is effectively a decrease of number density 
with the shot noise due to the randoms becoming dominant. 
This increased noise impacts the smaller scales more, and 
reduces the performance of the ES3-CNN. Since the ES3 reconstructed field 
is close to completely decorrelated on these scales, we don't see a similar degradation there.
At a minimum, one should use the number of randoms equal to the number of particles.

\subsection{Rotational invariance}
Rotational invariance is a fundamental property of large-scale structure; reconstruction should produce the same field for a rotated nonlinear field input. Our model produces the appropriately rotated output when given a rotated initial field, without training with rotated data or specifying rotational invariance in the network. Both HE18- and ES3-CNN can recognize both 180$^\circ$ and 90$^\circ$ rotations; the predictions of a grid rotated by 180$^\circ$ and 90$^\circ$ are on top of the prediction of an unrotated grid, preserving these types of rotation\footnote{In the process of testing rotational invariance, we first found that the ES3-CNN was not invariant under 90$^\circ$ rotations. This led us to discover a subtle bug in our ES3 reconstruction code, highlighting the power of invariance in detecting bugs.}. The difference in the loss between rotated and the unrotated fields is 0.01\%. In redshift space, rotation is preserved along the redshift direction ($z$ axis). The preservation of rotational symmetry adds more validity to our model. An extension for future work can be building rotational invariance in the network to reduce the number of parameters needed by the network.

\subsection{Effects of hyperparameters in the model}
\label{sec:hyper}

We parameterize the size of the network with the number of channels in the first two convolutional layers, $N_{1}$, with $N_{1}=32$ used in the original network by \citet{Mao21}. The next two convolutional layers have number of channels $N_2=2\times N_1$ and the next three convolutional layers after these have number of channels $N_3=2\times N_2$. We find that increasing $N_{1}$ to 48 only marginally improves the performance. We also test the batch sizes of $16^3$, $32^3$, $64^3$, and $128^3$. Smaller batch sizes appear to result in improved $G(k)$ and $r(k)$ measurements, although this saturates around $32^3$. An even smaller batch size also results in a biased propagator. These experiments motivate our fiducial choices of network architecture ($N_{1}=32$) and batch size of $32^3$. We defer a detailed exploration of the optimal hyperparameter settings to future work.

\subsection{Comparing with previous work}
While our model is based on the model proposed by \citet{Mao21}, 
using reconstructed field as input leads to a significant improvement over their results. 
As shown in Figure 5 in \citet{Mao21} as well as our Figure~\ref{fig:defaultcos}, when a network is trained with unreconstructed data, 
the performance is generally worse than reconstruction algorithms alone. 
The worse performance with training pre-reconstruction field is because, while the CNN can
learn local features well, it does not have access to large-scale information. 
Once the large-scale reconstruction is performed by the reconstruction algorithm, the CNN can continue on the smaller-scale reconstruction. This advantage of using the CNN to follow up on the reconstructed
data was also pointed out in \citet{Shallue22}, although the network architecture used was different.

The most direct comparison with our work is the previous work by \citet[][hereafter SE22]{Shallue22}, where they introduced the idea of applying a CNN to the reconstructed density field, instead of the nonlinear field itself. This work extends that work in two notable directions -- we explore the dependence of this approach on the underlying reconstruction technique, and the impact of shot noise on the performance. We also note that SE22 uses a different network architecture from ours, although the underlying motivation of using local information to reconstruct the field is the same.

\begin{figure}
    \centering
    \includegraphics[width=\columnwidth]{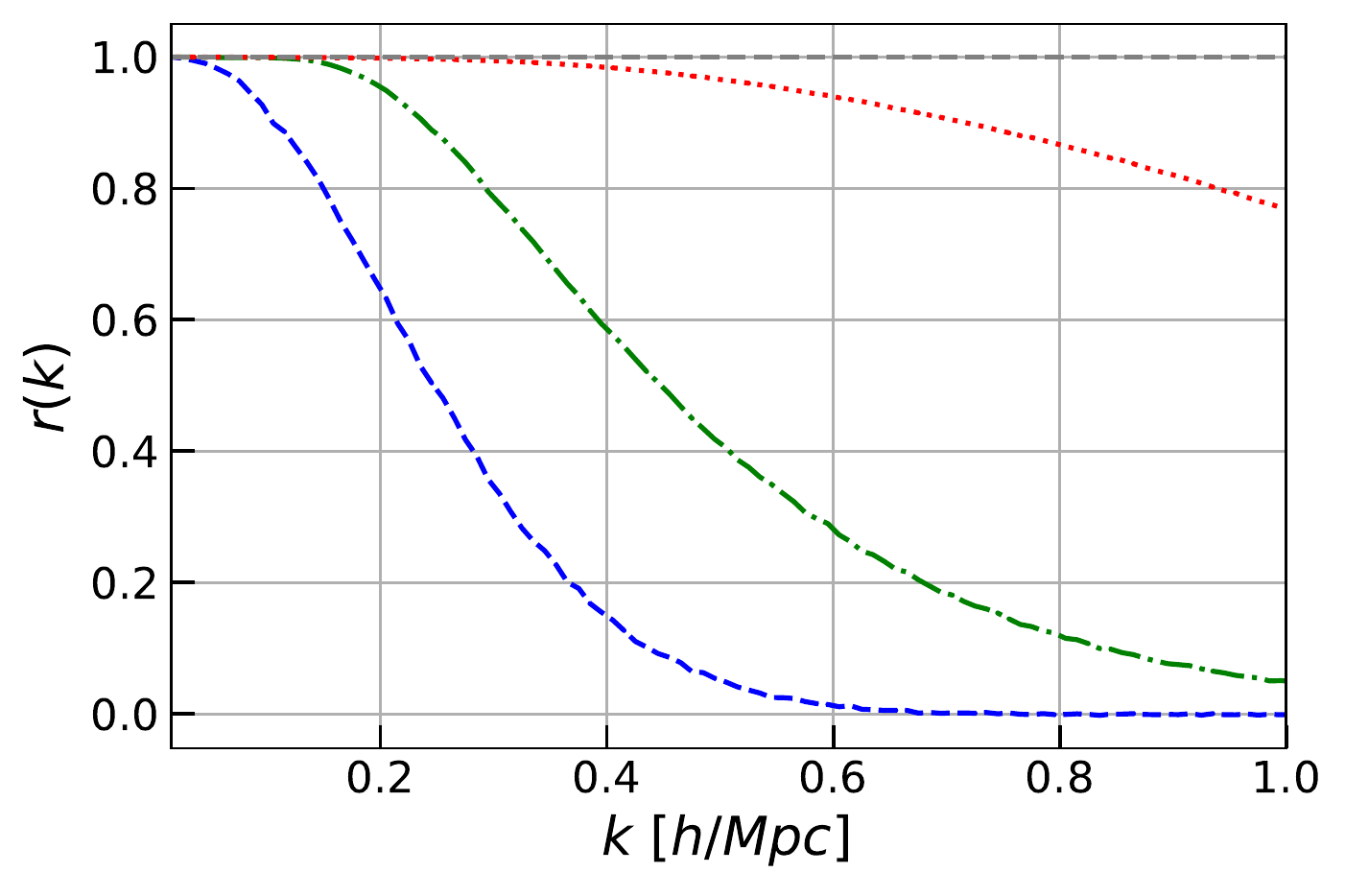}
    \caption{A model trained with $z=0.5$ real space data reconstructed by ES3 algorithm with our architecture to compare with the performance of \citet[][]{Shallue22} method (c.f. the left panel of Figure 2 in \citet{Shallue22}). One simulation is used in inference. Line styles follow those in \citet{Shallue22}. Red dotted line is our CNN with ES3 reconstruction, green dot-dashed line is ES3 reconstruction, and blue dashed line is pre-reconstruction. The performance is similar to that of \citet[][]{Shallue22} method. 
}
    \label{fig:compare_shallue}
\end{figure}

To better compare the approaches, we train our model on $z=0.5$ data snapshots; the results are in Figure~\ref{fig:compare_shallue}. We see very similar behavior -- the SE22 model reaches $r(k)=0.8$ at $k\sim0.75 h\Mpc^{-1}$, while our work reaches the same at $k \sim 0.93 h\Mpc^{-1}$. We however note that we did not attempt to optimize the SE22 architecture for these comparisons, nor did we try to match training hyperparameters, simulations etc. -- all of which could contribute to the differences. We also observe the same issue with the redshift space quadrupole, where the ES3-CNN model produces spurious power on large scales (but the HE18-CNN model does not). What these comparisons emphasize is the potential utility of augmenting machine-learning methods with more classical approaches.

We also note that our CNN method achieves comparable performance to the level of $\mathcal{O}(2)$ reconstruction by \citet[][]{Schmittfull17} in real space at $z=0$ (c.f. Figures 4 and 10 in \citet[][]{Schmittfull17}). This method iteratively computes the displacement field working to second order in the density field.
HE18 also considers the second order, but in the estimate of the displacement rather than density field, while ES3 does not consider any higher order corrections. We point out in \citet{reconpaper} that the second order displacement in HE18 makes little difference. In \citet{Schmittfull17}, the improvement of $\mathcal{O}(2)$ over $\mathcal{O}(1)$ is also marginal. So it is unlikely the higher order estimation that contributes to a better approximation to the initial condition in both our work and \citet{Schmittfull17}, but a much better use of information of the field by CNN and the iterative algorithm used in \citet{Schmittfull17}. It is worth investigating whether the CNN is learning similar information that is employed in \citet{Schmittfull17}.

\section{Summary and future work}\label{sect:summary}
We build an eight-layer CNN for cosmic density field reconstruction and train it with reconstructed fields produced by traditional algorithms. The result achieves significantly higher closeness to the initial condition than using reconstruction algorithms alone. The model can also be applied to other cosmologies and redshifts and still achieve much better results than reconstruction algorithms. It is also more effective at removing the residual redshift space distortions left by the standard methods. 

Our primary conclusions are
\begin{enumerate}
    \item The CNN models learn patterns in the field that are remarkably robust to changes in the underlying cosmology or the redshifts of the data it is trained on.
    \item Processing with the CNN largely erases the differences between the various reconstruction algorithms, including the effect of the smoothing scales adopted.
    \item The one difference between the two input reconstruction algorithms is in redshift space, where the ES3-CNN appears to introduce spurious large-scale power in the quadrupole.
    \item The CNN methods are quite sensitive to noise in the input field and their performance degrades significantly with increasing shot noise. This suggests that these algorithms are best suited for high number density samples. This observation might help inform the design of future surveys.
\end{enumerate}

There are a number of extensions possible of this work; we mention some of these here:
\begin{enumerate}
    \item Application to biased fields : the work here has focused on the matter density field. A direct extension would be to biased tracers.
    \item Imposing symmetries : we demonstrated that the network here could learn rotational invariance even if not explicitly trained to recognize it. It would be interesting to explore if future architectures could explicitly build in these symmetries.
    \item Using other reconstruction algorithms as input : as CNN enhanced ES3 and HE18 reconstructions achieve similar results, it would be interesting to test whether other newly proposed reconstruction algorithms, when coupled with our CNN, also achieve higher fidelity and at a similar level to this work.
    \item Fitting the reconstructed statistics: given the improved reconstruction of the linear power spectrum, one could explore the fitting of the two-point and higher order statistics.
\end{enumerate}

\section*{Acknowledgements}


We thank Daniel Eisenstein, Christopher Shallue, Sihao Cheng, Zachary Slepian, Jiamin Hou, Mehdi Rezaie, Marc Huertas-Company, John Lafferty, Farnik Nikakhtar, Sheridan Green, Carolina Cuesta-Lazaro, Chirag Modi, Benjamin Nachman, Cora Dvorkin, Georgios Valogiannis, and Jasjeet Sekhon for helpful discussions. This paper also benefited from a detailed set of comments from an anonymous referee. XC is supported by Future Investigators in NASA Earth and Space Science and Technology (FINESST) grant (award \#80NSSC21K2041). SG is supported by the National Science Foundation Graduate Research Fellowship Program under grant No. DGE-1752134. NP is supported in part by DOE DE-SC0017660.

\section*{Data Availability}

The code and trained CNN models developed in this work as well as statistics produced in analysis are available upon request. The \Quijote simulations used in this work are publicly available
with details in \citet{Navarro20}.
 



\bibliographystyle{mnras}
\bibliography{CNN_Recon_paper} 

\begin{thebibliography}{}
\makeatletter
\relax
\def\mn@urlcharsother{\let\do\@makeother \do\$\do\&\do\#\do\^\do\_\do\%\do\~}
\def\mn@doi{\begingroup\mn@urlcharsother \@ifnextchar [ {\mn@doi@}
  {\mn@doi@[]}}
\def\mn@doi@[#1]#2{\def\@tempa{#1}\ifx\@tempa\@empty \href
  {http://dx.doi.org/#2} {doi:#2}\else \href {http://dx.doi.org/#2} {#1}\fi
  \endgroup}
\def\mn@eprint#1#2{\mn@eprint@#1:#2::\@nil}
\def\mn@eprint@arXiv#1{\href {http://arxiv.org/abs/#1} {{\tt arXiv:#1}}}
\def\mn@eprint@dblp#1{\href {http://dblp.uni-trier.de/rec/bibtex/#1.xml}
  {dblp:#1}}
\def\mn@eprint@#1:#2:#3:#4\@nil{\def\@tempa {#1}\def\@tempb {#2}\def\@tempc
  {#3}\ifx \@tempc \@empty \let \@tempc \@tempb \let \@tempb \@tempa \fi \ifx
  \@tempb \@empty \def\@tempb {arXiv}\fi \@ifundefined
  {mn@eprint@\@tempb}{\@tempb:\@tempc}{\expandafter \expandafter \csname
  mn@eprint@\@tempb\endcsname \expandafter{\@tempc}}}

\bibitem[\protect\citeauthoryear{Aghanim et~al.,}{Aghanim
  et~al.}{2020}]{Planck18}
Aghanim N.,  et~al., 2020, \mn@doi [Astronomy & Astrophysics]
  {10.1051/0004-6361/201833910}, 641, A6

\bibitem[\protect\citeauthoryear{{Alcock} \& {Paczynski}}{{Alcock} \&
  {Paczynski}}{1979}]{Alcock79}
{Alcock} C.,  {Paczynski} B.,  1979, \mn@doi [\nat] {10.1038/281358a0}, \href
  {https://ui.adsabs.harvard.edu/abs/1979Natur.281..358A} {281, 358}

\bibitem[\protect\citeauthoryear{Anderson et~al.,}{Anderson
  et~al.}{2014}]{Anderson14b}
Anderson L.,  et~al., 2014, \mn@doi [Monthly Notices of the Royal Astronomical
  Society] {10.1093/mnras/stt2206}, 439, 83

\bibitem[\protect\citeauthoryear{Beutler et~al.,}{Beutler
  et~al.}{2016}]{Beutler17}
Beutler F.,  et~al., 2016, \mn@doi [Monthly Notices of the Royal Astronomical
  Society] {10.1093/mnras/stw2373}, 464, 3409

\bibitem[\protect\citeauthoryear{{Chen} \& {Padmanabhan}}{{Chen} \&
  {Padmanabhan}}{2023}]{reconpaper}
{Chen} X.,  {Padmanabhan} N.,  2023, In preperation

\bibitem[\protect\citeauthoryear{{Chen}, {Vlah}  \& {White}}{{Chen}
  et~al.}{2019}]{Chen19}
{Chen} S.-F.,  {Vlah} Z.,   {White} M.,  2019, \mn@doi [\jcap]
  {10.1088/1475-7516/2019/09/017}, \href
  {https://ui.adsabs.harvard.edu/abs/2019JCAP...09..017C} {2019, 017}

\bibitem[\protect\citeauthoryear{{Crocce} \& {Scoccimarro}}{{Crocce} \&
  {Scoccimarro}}{2008}]{Crocce08}
{Crocce} M.,  {Scoccimarro} R.,  2008, \mn@doi [\prd]
  {10.1103/PhysRevD.77.023533}, \href
  {https://ui.adsabs.harvard.edu/abs/2008PhRvD..77b3533C} {77, 023533}

\bibitem[\protect\citeauthoryear{{DESI Collaboration} et~al.,}{{DESI
  Collaboration} et~al.}{2016}]{DESI}
{DESI Collaboration} et~al., 2016, The DESI Experiment Part I:
  Science,Targeting, and Survey Design (\mn@eprint {arXiv} {1611.00036})

\bibitem[\protect\citeauthoryear{{Dai}, {Feng}  \& {Seljak}}{{Dai}
  et~al.}{2018}]{Dai18}
{Dai} B.,  {Feng} Y.,   {Seljak} U.,  2018, \mn@doi [\jcap]
  {10.1088/1475-7516/2018/11/009}, \href
  {https://ui.adsabs.harvard.edu/abs/2018JCAP...11..009D} {2018, 009}

\bibitem[\protect\citeauthoryear{{Eisenstein}, {Seo}, {Sirko}  \&
  {Spergel}}{{Eisenstein} et~al.}{2007a}]{Eisenstein07}
{Eisenstein} D.~J.,  {Seo} H.-J.,  {Sirko} E.,   {Spergel} D.~N.,  2007a,
  \mn@doi [\apj] {10.1086/518712}, \href
  {http://adsabs.harvard.edu/abs/2007ApJ...664..675E} {664, 675}

\bibitem[\protect\citeauthoryear{Eisenstein, Seo  \& White}{Eisenstein
  et~al.}{2007b}]{Eisenstein07b}
Eisenstein D.~J.,  Seo H.,   White M.,  2007b, \mn@doi [The Astrophysical
  Journal] {10.1086/518755}, 664, 660

\bibitem[\protect\citeauthoryear{{Etezad-Razavi}, {Abbasgholinejad},
  {Sotoudeh}, {Hassani}, {Raeisi}  \& {Baghram}}{{Etezad-Razavi}
  et~al.}{2021}]{Etezad-Razavi21}
{Etezad-Razavi} S.,  {Abbasgholinejad} E.,  {Sotoudeh} M.-H.,  {Hassani} F.,
  {Raeisi} S.,   {Baghram} S.,  2021, arXiv e-prints, \href
  {https://ui.adsabs.harvard.edu/abs/2021arXiv211214743E} {p. arXiv:2112.14743}

\bibitem[\protect\citeauthoryear{Fluri, Kacprzak, Refregier, Amara, Lucchi  \&
  Hofmann}{Fluri et~al.}{2018}]{Fluri18}
Fluri J.,  Kacprzak T.,  Refregier A.,  Amara A.,  Lucchi A.,   Hofmann T.,
  2018, \mn@doi [Phys. Rev. D] {10.1103/PhysRevD.98.123518}, 98, 123518

\bibitem[\protect\citeauthoryear{Fukushima \& Miyake}{Fukushima \&
  Miyake}{1982}]{Fukushima82}
Fukushima K.,  Miyake S.,  1982, in Amari S.-i.,  Arbib M.~A.,  eds,
  Competition and Cooperation in Neural Nets. Springer Berlin Heidelberg,
  Berlin, Heidelberg, pp 267--285

\bibitem[\protect\citeauthoryear{{Gil-Mar{\'\i}n} et~al.,}{{Gil-Mar{\'\i}n}
  et~al.}{2020}]{Gil20}
{Gil-Mar{\'\i}n} H.,  et~al., 2020, \mn@doi [\mnras] {10.1093/mnras/staa2455},
  \href {https://ui.adsabs.harvard.edu/abs/2020MNRAS.498.2492G} {498, 2492}

\bibitem[\protect\citeauthoryear{Gupta, Matilla, Hsu  \& Haiman}{Gupta
  et~al.}{2018}]{Gupta18}
Gupta A.,  Matilla J. M.~Z.,  Hsu D.,   Haiman Z.,  2018, \mn@doi [Phys. Rev.
  D] {10.1103/PhysRevD.97.103515}, 97, 103515

\bibitem[\protect\citeauthoryear{{Hada} \& {Eisenstein}}{{Hada} \&
  {Eisenstein}}{2018}]{Hada18}
{Hada} R.,  {Eisenstein} D.~J.,  2018, \mn@doi [\mnras]
  {10.1093/mnras/sty1203}, \href
  {http://adsabs.harvard.edu/abs/2018MNRAS.478.1866H} {478, 1866}

\bibitem[\protect\citeauthoryear{Ho, Rau, Ntampaka, Farahi, Trac  \&
  P{\'o}czos}{Ho et~al.}{2019}]{Ho19}
Ho M.,  Rau M.~M.,  Ntampaka M.,  Farahi A.,  Trac H.,   P{\'o}czos B.,  2019,
  \mn@doi [The Astrophysical Journal] {10.3847/1538-4357/ab4f82}, 887, 25

\bibitem[\protect\citeauthoryear{{Hockney} \& {Eastwood}}{{Hockney} \&
  {Eastwood}}{1988}]{Hockney88}
{Hockney} R.~W.,  {Eastwood} J.~W.,  1988, {Computer simulation using
  particles}.
Bristol: Hilger

\bibitem[\protect\citeauthoryear{{Huertas-Company} \&
  {Lanusse}}{{Huertas-Company} \& {Lanusse}}{2022}]{Company22}
{Huertas-Company} M.,  {Lanusse} F.,  2022, arXiv e-prints, \href
  {https://ui.adsabs.harvard.edu/abs/2022arXiv221001813H} {p. arXiv:2210.01813}

\bibitem[\protect\citeauthoryear{Jasche \& Wandelt}{Jasche \&
  Wandelt}{2013}]{Jasche13}
Jasche J.,  Wandelt B.~D.,  2013, \mn@doi [Monthly Notices of the Royal
  Astronomical Society] {10.1093/mnras/stt449}, 432, 894

\bibitem[\protect\citeauthoryear{{Kaiser}}{{Kaiser}}{1987}]{Kaiser87}
{Kaiser} N.,  1987, \mn@doi [\mnras] {10.1093/mnras/227.1.1}, \href
  {http://adsabs.harvard.edu/abs/1987MNRAS.227....1K} {227, 1}

\bibitem[\protect\citeauthoryear{{Kingma} \& {Ba}}{{Kingma} \&
  {Ba}}{2014}]{Kingma14}
{Kingma} D.~P.,  {Ba} J.,  2014, arXiv e-prints, \href
  {https://ui.adsabs.harvard.edu/abs/2014arXiv1412.6980K} {p. arXiv:1412.6980}

\bibitem[\protect\citeauthoryear{{Kitaura}}{{Kitaura}}{2013}]{Kitaura13}
{Kitaura} F.~S.,  2013, \mn@doi [\mnras] {10.1093/mnrasl/sls029}, \href
  {https://ui.adsabs.harvard.edu/abs/2013MNRAS.429L..84K} {429, L84}

\bibitem[\protect\citeauthoryear{Krizhevsky, Sutskever  \& Hinton}{Krizhevsky
  et~al.}{2012}]{Krizhevsky12}
Krizhevsky A.,  Sutskever I.,   Hinton G.~E.,  2012, in Pereira F.,  Burges C.,
   Bottou L.,   Weinberger K.,  eds, ~ Vol. 25, Advances in Neural Information
  Processing Systems. Curran Associates, Inc., \url
  {https://proceedings.neurips.cc/paper/2012/file/c399862d3b9d6b76c8436e924a68c45b-Paper.pdf}

\bibitem[\protect\citeauthoryear{Laureijs et~al.,}{Laureijs
  et~al.}{2011}]{Laureijs11}
Laureijs R.,  et~al., 2011, Euclid Definition Study Report (\mn@eprint {arXiv}
  {1110.3193})

\bibitem[\protect\citeauthoryear{LeCun, Haffner, Bottou  \& Bengio}{LeCun
  et~al.}{1999}]{LeCun99}
LeCun Y.,  Haffner P.,  Bottou L.,   Bengio Y.,  1999, Object Recognition with
  Gradient-Based Learning.
Springer Berlin Heidelberg, Berlin, Heidelberg, pp 319--345,
  \mn@doi{10.1007/3-540-46805-6_19}, \url
  {https://doi.org/10.1007/3-540-46805-6_19}

\bibitem[\protect\citeauthoryear{{Levy}, {Mohayaee}  \& {von Hausegger}}{{Levy}
  et~al.}{2021}]{Levy21}
{Levy} B.,  {Mohayaee} R.,   {von Hausegger} S.,  2021, \mn@doi [\mnras]
  {10.1093/mnras/stab1676}, \href
  {https://ui.adsabs.harvard.edu/abs/2021MNRAS.506.1165L} {506, 1165}

\bibitem[\protect\citeauthoryear{{Lucie-Smith}, {Peiris}, {Pontzen}, {Nord}  \&
  {Thiyagalingam}}{{Lucie-Smith} et~al.}{2020}]{Lucie-Smith20}
{Lucie-Smith} L.,  {Peiris} H.~V.,  {Pontzen} A.,  {Nord} B.,   {Thiyagalingam}
  J.,  2020, arXiv e-prints, \href
  {https://ui.adsabs.harvard.edu/abs/2020arXiv201110577L} {p. arXiv:2011.10577}

\bibitem[\protect\citeauthoryear{{Mao}, {Wang}, {Li}, {Cai}, {Falck},
  {Neyrinck}  \& {Szalay}}{{Mao} et~al.}{2021}]{Mao21}
{Mao} T.-X.,  {Wang} J.,  {Li} B.,  {Cai} Y.-C.,  {Falck} B.,  {Neyrinck} M.,
  {Szalay} A.,  2021, \mn@doi [\mnras] {10.1093/mnras/staa3741}, \href
  {https://ui.adsabs.harvard.edu/abs/2021MNRAS.501.1499M} {501, 1499}

\bibitem[\protect\citeauthoryear{{Mathuriya} et~al.,}{{Mathuriya}
  et~al.}{2018}]{Mathuriya18}
{Mathuriya} A.,  et~al., 2018, arXiv e-prints, \href
  {https://ui.adsabs.harvard.edu/abs/2018arXiv180804728M} {p. arXiv:1808.04728}

\bibitem[\protect\citeauthoryear{{Meiksin}, {White}  \& {Peacock}}{{Meiksin}
  et~al.}{1999}]{Meiksin99}
{Meiksin} A.,  {White} M.,   {Peacock} J.~A.,  1999, \mn@doi [\mnras]
  {10.1046/j.1365-8711.1999.02369.x}, \href
  {https://ui.adsabs.harvard.edu/abs/1999MNRAS.304..851M} {304, 851}

\bibitem[\protect\citeauthoryear{Modi, Feng  \& Seljak}{Modi
  et~al.}{2018}]{Modi18}
Modi C.,  Feng Y.,   Seljak U.,  2018, \mn@doi [Journal of Cosmology and
  Astroparticle Physics] {10.1088/1475-7516/2018/10/028}, 2018, 028

\bibitem[\protect\citeauthoryear{Mustafa, Bard, Bhimji, Luki{\'c}, Al-Rfou  \&
  Kratochvil}{Mustafa et~al.}{2019}]{Mustafa19}
Mustafa M.,  Bard D.,  Bhimji W.,  Luki{\'c} Z.,  Al-Rfou R.,   Kratochvil
  J.~M.,  2019, \mn@doi [Computational Astrophysics and Cosmology]
  {10.1186/s40668-019-0029-9}, 6, 1

\bibitem[\protect\citeauthoryear{{Nikakhtar}, {Sheth}  \& {Zehavi}}{{Nikakhtar}
  et~al.}{2021}]{Nikakhtar21}
{Nikakhtar} F.,  {Sheth} R.~K.,   {Zehavi} I.,  2021, \mn@doi [\prd]
  {10.1103/PhysRevD.104.043530}, \href
  {https://ui.adsabs.harvard.edu/abs/2021PhRvD.104d3530N} {104, 043530}

\bibitem[\protect\citeauthoryear{{Nikakhtar}, {Sheth}, {L{\'e}vy}  \&
  {Mohayaee}}{{Nikakhtar} et~al.}{2022}]{Nikakhtar22}
{Nikakhtar} F.,  {Sheth} R.~K.,  {L{\'e}vy} B.,   {Mohayaee} R.,  2022, arXiv
  e-prints, \href {https://ui.adsabs.harvard.edu/abs/2022arXiv220301868N} {p.
  arXiv:2203.01868}

\bibitem[\protect\citeauthoryear{Ntampaka et~al.,}{Ntampaka
  et~al.}{2019}]{Ntampaka19}
Ntampaka M.,  et~al., 2019, \mn@doi [The Astrophysical Journal]
  {10.3847/1538-4357/ab14eb}, 876, 82

\bibitem[\protect\citeauthoryear{{Ntampaka}, {Eisenstein}, {Yuan}  \&
  {Garrison}}{{Ntampaka} et~al.}{2020}]{Ntampaka20}
{Ntampaka} M.,  {Eisenstein} D.~J.,  {Yuan} S.,   {Garrison} L.~H.,  2020,
  \mn@doi [\apj] {10.3847/1538-4357/ab5f5e}, \href
  {https://ui.adsabs.harvard.edu/abs/2020ApJ...889..151N} {889, 151}

\bibitem[\protect\citeauthoryear{{Obuljen}, {Villaescusa-Navarro}, {Castorina}
  \& {Viel}}{{Obuljen} et~al.}{2017}]{Obuljen17}
{Obuljen} A.,  {Villaescusa-Navarro} F.,  {Castorina} E.,   {Viel} M.,  2017,
  \mn@doi [Journal of Cosmology and Astro-Particle Physics]
  {10.1088/1475-7516/2017/09/012}, \href
  {https://ui.adsabs.harvard.edu/\#abs/2017JCAP...09..012O} {2017, 012}

\bibitem[\protect\citeauthoryear{Padmanabhan \& White}{Padmanabhan \&
  White}{2009}]{padmanabhan09b}
Padmanabhan N.,  White M.,  2009, \mn@doi [Physical Review D]
  {10.1103/physrevd.80.063508}, 80

\bibitem[\protect\citeauthoryear{{Padmanabhan}, {Xu}, {Eisenstein}, {Scalzo},
  {Cuesta}, {Mehta}  \& {Kazin}}{{Padmanabhan} et~al.}{2012}]{Padmanabhan12}
{Padmanabhan} N.,  {Xu} X.,  {Eisenstein} D.~J.,  {Scalzo} R.,  {Cuesta} A.~J.,
   {Mehta} K.~T.,   {Kazin} E.,  2012, \mn@doi [\mnras]
  {10.1111/j.1365-2966.2012.21888.x}, \href
  {http://adsabs.harvard.edu/abs/2012MNRAS.427.2132P} {427, 2132}

\bibitem[\protect\citeauthoryear{{Ravanbakhsh}, {Oliva}, {Fromenteau}, {Price},
  {Ho}, {Schneider}  \& {Poczos}}{{Ravanbakhsh} et~al.}{2017}]{Ravanbakhsh17}
{Ravanbakhsh} S.,  {Oliva} J.,  {Fromenteau} S.,  {Price} L.~C.,  {Ho} S.,
  {Schneider} J.,   {Poczos} B.,  2017, arXiv e-prints, \href
  {https://ui.adsabs.harvard.edu/abs/2017arXiv171102033R} {p. arXiv:1711.02033}

\bibitem[\protect\citeauthoryear{Ribli, Pataki, Zorrilla~Matilla, Hsu, Haiman
  \& Csabai}{Ribli et~al.}{2019}]{Ribli19}
Ribli D.,  Pataki B.~{\'A}.,  Zorrilla~Matilla J.~M.,  Hsu D.,  Haiman Z.,
  Csabai I.,  2019, \mn@doi [Monthly Notices of the Royal Astronomical Society]
  {10.1093/mnras/stz2610}, 490, 1843

\bibitem[\protect\citeauthoryear{{Ross} et~al.,}{{Ross} et~al.}{2017}]{Ross17}
{Ross} A.~J.,  et~al., 2017, \mn@doi [\mnras] {10.1093/mnras/stw2372}, \href
  {https://ui.adsabs.harvard.edu/abs/2017MNRAS.464.1168R} {464, 1168}

\bibitem[\protect\citeauthoryear{{Schmittfull}, {Feng}, {Beutler}, {Sherwin}
  \& {Chu}}{{Schmittfull} et~al.}{2015}]{Schmittfull15}
{Schmittfull} M.,  {Feng} Y.,  {Beutler} F.,  {Sherwin} B.,   {Chu} M.~Y.,
  2015, \mn@doi [\prd] {10.1103/PhysRevD.92.123522}, \href
  {https://ui.adsabs.harvard.edu/\#abs/2015PhRvD..92l3522S} {92, 123522}

\bibitem[\protect\citeauthoryear{{Schmittfull}, {Baldauf}  \&
  {Zaldarriaga}}{{Schmittfull} et~al.}{2017}]{Schmittfull17}
{Schmittfull} M.,  {Baldauf} T.,   {Zaldarriaga} M.,  2017, \mn@doi [\prd]
  {10.1103/PhysRevD.96.023505}, \href
  {https://ui.adsabs.harvard.edu/\#abs/2017PhRvD..96b3505S} {96, 023505}

\bibitem[\protect\citeauthoryear{{Seljak}, {Aslanyan}, {Feng}  \&
  {Modi}}{{Seljak} et~al.}{2017}]{Seljak17}
{Seljak} U.,  {Aslanyan} G.,  {Feng} Y.,   {Modi} C.,  2017, \mn@doi [\jcap]
  {10.1088/1475-7516/2017/12/009}, \href
  {https://ui.adsabs.harvard.edu/abs/2017JCAP...12..009S} {2017, 009}

\bibitem[\protect\citeauthoryear{Seo \& Hirata}{Seo \& Hirata}{2016}]{Seo16}
Seo H.-J.,  Hirata C.~M.,  2016, \mn@doi [Monthly Notices of the Royal
  Astronomical Society] {10.1093/mnras/stv2806}, 456, 3142

\bibitem[\protect\citeauthoryear{Seo et~al.,}{Seo et~al.}{2010}]{Seo10}
Seo H.-J.,  et~al., 2010, \mn@doi [The Astrophysical Journal]
  {10.1088/0004-637x/720/2/1650}, 720, 1650

\bibitem[\protect\citeauthoryear{Seo, Beutler, Ross  \& Saito}{Seo
  et~al.}{2016}]{Seo16b}
Seo H.-J.,  Beutler F.,  Ross A.~J.,   Saito S.,  2016, \mn@doi [Monthly
  Notices of the Royal Astronomical Society] {10.1093/mnras/stw1138}, 460, 2453

\bibitem[\protect\citeauthoryear{{Seo}, {Ota}, {Schmittfull}, {Saito}  \&
  {Beutler}}{{Seo} et~al.}{2022}]{Seo22}
{Seo} H.-J.,  {Ota} A.,  {Schmittfull} M.,  {Saito} S.,   {Beutler} F.,  2022,
  \mn@doi [\mnras] {10.1093/mnras/stac082}, \href
  {https://ui.adsabs.harvard.edu/abs/2022MNRAS.511.1557S} {511, 1557}

\bibitem[\protect\citeauthoryear{{Shallue} \& {Eisenstein}}{{Shallue} \&
  {Eisenstein}}{2022}]{Shallue22}
{Shallue} C.~J.,  {Eisenstein} D.~J.,  2022, arXiv e-prints, \href
  {https://ui.adsabs.harvard.edu/abs/2022arXiv220712511S} {p. arXiv:2207.12511}

\bibitem[\protect\citeauthoryear{{Simonyan} \& {Zisserman}}{{Simonyan} \&
  {Zisserman}}{2014}]{Simonyan14}
{Simonyan} K.,  {Zisserman} A.,  2014, arXiv e-prints, \href
  {https://ui.adsabs.harvard.edu/abs/2014arXiv1409.1556S} {p. arXiv:1409.1556}

\bibitem[\protect\citeauthoryear{{Spergel} et~al.,}{{Spergel}
  et~al.}{2013}]{Spergel13}
{Spergel} D.,  et~al., 2013, arXiv e-prints, \href
  {https://ui.adsabs.harvard.edu/\#abs/2013arXiv1305.5422S} {p.
  arXiv:1305.5422}

\bibitem[\protect\citeauthoryear{Tassev \& Zaldarriaga}{Tassev \&
  Zaldarriaga}{2012}]{Tassev12}
Tassev S.,  Zaldarriaga M.,  2012, \mn@doi [Journal of Cosmology and
  Astroparticle Physics] {10.1088/1475-7516/2012/10/006}, 2012, 006

\bibitem[\protect\citeauthoryear{{Vargas-Maga{\~n}a}
  et~al.,}{{Vargas-Maga{\~n}a} et~al.}{2018}]{Vargas18}
{Vargas-Maga{\~n}a} M.,  et~al., 2018, \mn@doi [\mnras] {10.1093/mnras/sty571},
  \href {https://ui.adsabs.harvard.edu/abs/2018MNRAS.477.1153V} {477, 1153}

\bibitem[\protect\citeauthoryear{{Villaescusa-Navarro}
  et~al.,}{{Villaescusa-Navarro} et~al.}{2020}]{Navarro20}
{Villaescusa-Navarro} F.,  et~al., 2020, \mn@doi [\apjs]
  {10.3847/1538-4365/ab9d82}, \href
  {https://ui.adsabs.harvard.edu/abs/2020ApJS..250....2V} {250, 2}

\bibitem[\protect\citeauthoryear{{Wang}, {Mo}, {Yang}, {Jing}  \& {Lin}}{{Wang}
  et~al.}{2014}]{Wang14}
{Wang} H.,  {Mo} H.~J.,  {Yang} X.,  {Jing} Y.~P.,   {Lin} W.~P.,  2014,
  \mn@doi [\apj] {10.1088/0004-637X/794/1/94}, \href
  {https://ui.adsabs.harvard.edu/abs/2014ApJ...794...94W} {794, 94}

\bibitem[\protect\citeauthoryear{{White}}{{White}}{2010}]{white10}
{White} M.,  2010, arXiv e-prints, \href
  {https://ui.adsabs.harvard.edu/abs/2010arXiv1004.0250W} {p. arXiv:1004.0250}

\bibitem[\protect\citeauthoryear{{White}}{{White}}{2015}]{White15}
{White} M.,  2015, \mn@doi [\mnras] {10.1093/mnras/stv842}, \href
  {https://ui.adsabs.harvard.edu/abs/2015MNRAS.450.3822W} {450, 3822}

\bibitem[\protect\citeauthoryear{{Xu}, {Cuesta}, {Padmanabhan}, {Eisenstein}
  \& {McBride}}{{Xu} et~al.}{2013}]{Xu13}
{Xu} X.,  {Cuesta} A.~J.,  {Padmanabhan} N.,  {Eisenstein} D.~J.,   {McBride}
  C.~K.,  2013, \mn@doi [\mnras] {10.1093/mnras/stt379}, \href
  {http://adsabs.harvard.edu/abs/2013MNRAS.431.2834X} {431, 2834}

\bibitem[\protect\citeauthoryear{{Zel'Dovich}}{{Zel'Dovich}}{1970}]{Zeldovich70}
{Zel'Dovich} Y.~B.,  1970, \aap, \href
  {https://ui.adsabs.harvard.edu/abs/1970A&A.....5...84Z} {500, 13}

\makeatother
\end{thebibliography}




\appendix

\section{A toy model for shot noise}\label{appx:Wiener} 
In this model, we assume that the input to the network is a sum of the linear density field and the shot noise, which is assumed to be uncorrelated with the density field: $\delta_{\rm in}=\delta_{\rm L}+\epsilon$. The target field is the linear density field, $\delta_{L}$.

We further assume that the input field has the property that the individual $k$-modes are independent of each other. Then by Parseval's theorem, the loss function has the same expression in Fourier and configuration spaces. Our loss function here is the same as previously used, $L=\sum_{i}\left[\hat{\delta}({\boldsymbol{r}_i)}-\delta_{\rm L}(\boldsymbol{r}_i)\right]^2$, where $\hat{\delta}({\boldsymbol{r}_i)}$ is the estimated field. The Fourier space loss function in term takes the form:
\begin{equation}
L=\sum_{i}\left[\hat{\delta}({\boldsymbol{k}_i)}-\delta_{\rm L}(\boldsymbol{k}_i)\right]^2. 
\end{equation}
Since the $k$-modes are assumed to be independent, we write the estimated field as a linear function of the input (Equation~\ref{eqn:delta_hat}). The loss is then
\begin{equation}
    L(k)=\left<[W(k)(\delta_{\rm L}(k)+\epsilon)-\delta_{\rm L}(k)]^2\right>.
\end{equation}

Taking the first derivative of the above and setting it to 0, we arrive at
\begin{equation}
    \frac{\partial L(k)}{\partial W(k)}=\left<2\left[W(k)(\delta_{\rm L}(k)+\epsilon)-\delta_{\rm L}(k)\right](\delta_{\rm L}(k)+\epsilon)\right>=0,
\end{equation}
which leads to 
\begin{equation}
    W(k)\left[P_{\rm L}(k)+\frac{1}{\Bar{n}}\right]-P_{\rm L}(k)=0,
\end{equation}
where $\langle(\delta_{\rm L}(k)+\epsilon)^2\rangle=P_{\rm L}(k)+\frac{1}{\Bar{n}}$, since we assume that shot noise is uncorrelated. The solution to the above equation is the Wiener filter
\begin{equation}
    W(k)=\frac{P_{\rm L}(k)}{P_{\rm L}(k)+\frac{1}{\Bar{n}}}.
\end{equation}



\bsp	
\label{lastpage}
\end{document}